\documentclass[preprint,nofootinbib,floatfix,a4paper,prd,superscriptaddress]{revtex4-1}   	
\usepackage{geometry}                		
\usepackage[colorlinks,citecolor=blue]{hyperref}
\usepackage{amsmath}
\usepackage{mathrsfs}
\usepackage{bbm}
\usepackage{amsfonts}
\usepackage{amssymb}
\usepackage{latexsym}
\usepackage{graphicx}
\usepackage[english]{babel}
\usepackage{multirow}
\usepackage{float}
\usepackage{url}
\usepackage{slashed}

\newcommand{\be}{\begin{equation}}
\newcommand{\ee}{\end{equation}}
\newcommand{\ba}{\begin{array}}
\newcommand{\ea}{\end{array}}
\newcommand{\bea}{\begin{eqnarray}}
\newcommand{\eea}{\end{eqnarray}}

\usepackage[utf8]{inputenc}
\usepackage[T1]{fontenc}
\usepackage{CormorantGaramond}

\def  \bcen   {\begin{center}}
\def  \ecen   {\end{center}}
\def  \beq    {\begin{equation}}
\def  \eeq    {\end{equation}}

\def\la   {\lambda}

\def\lee { \left( }
\def\rii { \right) }

\def\to {\rightarrow}

\begin{document}

\title{Charged Lepton Flavor Violating Radiative Decays $l_i \to l_j \gamma$ in G2HDM}

\author{Van Que Tran}
\email{vqtran@sjtu.edu.cn}
\affiliation{\small Tsung Dao Lee Institute $\&$ School of Physics and Astronomy, Shanghai Jiao Tong University, Shanghai 200240, China}
\affiliation{\small Faculty of Fundamental Sciences, PHENIKAA University, Yen Nghia, Ha Dong, Hanoi 12116, Vietnam}

\author{Tzu-Chiang Yuan}
\email{tcyuan@phys.sinica.edu.tw}
\affiliation{\small Institute of Physics, Academia Sinica, Nangang, Taipei 11529, Taiwan}


\begin{abstract}

We compute the electromagnetic form factors of the $l_i l_j \gamma$ vertex
at one-loop level in the minimal G2HDM which has a sub-GeV vector dark matter candidate. 
The results are applied to the radiative decay rates for the charged lepton flavor violating processes $l_i \to l_j \gamma$, 
and the anomalous magnetic dipole moment and the electric dipole moment of the charged lepton. 
To numerically compute the branching ratio
for $\mu \to e \gamma$ and compare with the latest experimental limit from MEG,
we adapt our previous parameter space scan that is consistent with the relic density
and constraints from direct searches of dark matter, 
$W$ and $Z$ mass measurements,
as well as the LHC Higgs signal strengths and invisible width. 
While the extra contributions are at least an order of magnitude smaller than required to explain the $\sim 4.2 \sigma$ discrepancy 
in the muon anomaly,
the existing MEG limit imposes stringent constraint on the parameter space.
The remaining viable parameter space can be further probed by the MEG II sensitivity for $\mu \to e \gamma$ as well as from 
the direct searches of sub-GeV dark matter
in foreseeable future.
Higher loop contributions may be significant to resolve the discrepancy in the muon anomaly and generate a non-vanishing electric dipole moments for the 
standard model quarks and leptons in G2HDM.

\end{abstract}

\maketitle


\section{Introduction}
\label{Intro}

Charged lepton flavor violating processes like $\mu \to e \gamma$, $\mu \to 3e$ or $\mu-e$ conversion in nuclei, {\it etc.} 
without any neutrino in the final states are absent at tree level in the standard model (SM) of particle physics. However they are 
not strictly forbidden by symmetry and can be induced by one-loop diagram with the $W$ boson exchange. 
Thus their branching ratios are vanishingly small as they are proportional to the neutrino masses~\cite{Petcov:1976ff,Cheng:1976uq,Wilczek:1977wb}. The most stringent experimental constraint 
is for $\mu \to e \gamma$, with the following limit on its branching ratio published in 2016 by the MEG collaboration~\cite{MEG:2016leq},
\beq
\mathcal B (\mu^+ \to e^+ \gamma)  < 4.2 \times 10^{-13} \; (90\%{\rm C.L.}) 
\eeq
and the projected future sensitivity is expected to improve about an order of magnitude
$\sim 6 \times 10^{-14}$ by MEG II~\cite{Meucci:2022qbh}. 
For reviews on the charged lepton flavor violation, see for example Refs.~\cite{Kuno:1999jp,deGouvea:2013zba,Lindner:2016bgg,Bernstein:2013hba}.

The process $\mu \to e \gamma$ (or in general $l_i \to l_j \gamma$ with the Latin indices $i,j,k=1,2,3$ labeling the generation (or flavor) here and henceforth) 
has been widely studied beyond the standard model (bSM)~\cite{Kuno:1999jp,deGouvea:2013zba,Lindner:2016bgg,Hung:2015hra,Hung:2017voe} with predictions on the branching ratios that are more reachable 
experimentally than the SM one in foreseeable future. Here we will study this process in the context of minimal 
gauged two-Higgs-doublet model (G2HDM)~\cite{Ramos:2021omo,Ramos:2021txu,Tran:2022yrh} which has a hidden SM-like dark sector of 
$SU(2)_H \times U(1)_X$ with a sub-GeV $\mathcal W^{\prime (p,m)}$ dark matter candidate. The stability of the dark matter in the model is due to a hidden $h$-parity which 
emerges naturally without introducing it on {\it ad hoc} basis. 
Under the $h$-parity, all the SM particles and extra neutral gauge bosons are even while other new particles in G2HDM are odd.

The new contributions to the one-loop process $l_i \to l_j \gamma$ in G2HDM
involve the new gauge or Yukawa couplings between a $h$-parity odd 
particle like the dark gauge boson $\mathcal W^{\prime (p,m)}$, complex scalar $\mathcal D$ or charged Higgs $\mathcal H^\pm$ 
couple to another $h$-parity odd heavy hidden leptons $l^H_k$ or $\nu^H_k$
and the external SM charged leptons $l_i$ and $l_j$. These new couplings are in general off-diagonal in the generation space and 
hence can give rise to  $l_i \to l_j$ transition with $i > j$ (in particular muon $\to$ electron) at one-loop. 
While the contribution from the dark charged Higgs $\mathcal H^\pm$ is suppressed by the neutrino masses like the SM $W^\pm$, the other new contributions 
are not and therefore can give rise to a branching ratio that is more accessible experimentally. Turning the argument around, one can use the present and future experimental limits on the charged lepton violating processes 
to constrain our model parameters in G2HDM.

As a byproduct of our computation of the form factors for $l_i \to l_j \gamma$, we can also
extract the anomalous magnetic dipole moment $a_{l_i}$ and the electric dipole moment $d_{l_i}$ easily by setting $i=j$ in our results. 
The muon anomalous magnetic dipole moment, 
\beq
a_\mu \equiv \left( \frac{g_\mu - 2}{2}  \right),
\eeq
where $g_\mu$ is the  $g-$factor of the muon, is the most precise measured quantity in SM, with a value
measured at the Brookhaven National Laboratory (BNL) E821 experiment (1997--2001)~\cite{Muong-2:2002wip,Muong-2:2004fok,Muong-2:2006rrc},
\beq
a_\mu({\rm BNL}) = (11 \, 659 \, 208.9 \pm   5.4_{\rm stat} \pm 3.3_{\rm sys} ) \times 10^{-10} \; .
\eeq
Recently, the Fermilab (FNAL) Muon $g-2$ Collaboration, based on the analysis of data set from Run 1 and Run 2, 
announced the first result on the measurement~\cite{Muong-2:2021ojo} 
\beq
a_\mu ({\rm FNAL}) = (11 \, 6 59 \, 204.0 \pm 5.4) \times 10^{-10} \; .
\eeq
The average value of $a_\mu$ from the two experiments is given by~\cite{Muong-2:2021ojo} 
\beq
a_\mu ({\rm BNL+FNAL}) = (11 \, 6 59 \, 206.1 \pm 4.1) \times 10^{-10} \; .
\eeq
For recent reviews of the muon $g-2$, see for example Refs.~\cite{Lindner:2016bgg,Aoyama:2020ynm,Keshavarzi:2021eqa}. 
The recommended value for the SM prediction of the muon $g-2$ is~\cite{Aoyama:2020ynm}
\beq
a^{\rm SM}_\mu = ( 11 \, 6 5 9 \, 181.0 \pm 4. 3 )\times 10^{-10} \; .
\eeq
Hence the discrepancy between the experimental and theoretical values amounts to~\cite{Muong-2:2021ojo} 
\beq
\Delta a_\mu \equiv a_\mu({\rm BNL+FNAL})-a_\mu^{\rm SM} = (25.1 \pm5.9) \times 10^{-10} \; ,
\eeq
which implies a significance at the 4.2$\sigma$ level, slightly under the standard criterion of 5$\sigma$ to claim a discovery. Nevertheless, 
this discrepancy is as large as the SM electroweak contribution to the muon $g-2$~\cite{Aoyama:2020ynm}, 
\beq
a_\mu^{\rm EW} = (15.4 \pm 0.1) \times 10^{-10} \; , 
\eeq
which provides strong hints of bSM physics around the electroweak scale 
be responsible for it. Future goal of the ongoing efforts at FNAL~\cite{Muong-2:2021ojo} is to further reduce the existing uncertainty in the muon anomaly measurement by a factor of 1/4.

For the electric dipole moment of the SM charged leptons, we 
show that they vanish identically at one-loop in minimal G2HDM 
due to the lack of CP violating phases in the products of related complex couplings as well as vanishing combinations of loop integrals.  
Similar conclusions can be obtained for the SM quarks.
Higher loop contributions are needed to anticipate to achieve a nonzero electric dipole moments
for the SM fermions in minimal G2HDM. We will not address this issue in this work.

Current experimental status of $a_\mu$, $\mathcal B ( \mu^+ \to e^+ \gamma)$ and $\vert d_{e,\mu} / e\vert$ are summarized in Table~\ref{observables}.

\begin{table}[htbp!]
\resizebox{\textwidth}{!}{
\begin{tabular}{|c|c|c|c|c|c|c|}
\hline
Observable & Experimental Result/Limit & Future Goal  \\
\hline\hline
$a_\mu({\rm BNL})$ & $(11 \, 659 \, 208.9 \pm   5.4_{\rm stat} \pm 3.3_{\rm sys} ) \times 10^{-10}$~\cite{Muong-2:2002wip,Muong-2:2004fok,Muong-2:2006rrc} & --  \\
$a_\mu ({\rm FNAL})$ & $(11 \, 6 59 \, 204.0 \pm 5.4) \times 10^{-10}$~\cite{Muong-2:2021ojo} &Uncertainty $\sim$1/4 of BNL  \\
$a_\mu ({\rm BNL+FNAL})$ & $(11 \, 6 59 \, 206.1 \pm 4.1) \times 10^{-10}$~\cite{Muong-2:2021ojo} &Uncertainty $\sim$1/4 of BNL \\
\hline
$\mathcal B (\mu^+ \to e^+ \gamma$) ({\rm MEG}) &$< 4.2 \times 10^{-13}  (90\%{\rm C.L.})$~\cite{MEG:2016leq} &$\sim 6 \times 10^{-14}$ (MEG II~\cite{Meucci:2022qbh}) \\
$\vert \frac{d_\mu}{e} \vert$ [cm]  & $< 1.8 \times 10^{-19} (95\%{\rm C.L.})$~\cite{Muong-2:2008ebm}  & $\sim 6 \times 10^{-23}$ (PSI~\cite{Adelmann:2021udj}) \\
$\vert \frac{d_e}{e} \vert$ [cm] ({\rm ACME}) &$ < 1.1\times 10^{-29} (90\%{\rm C.L.})$~\cite{ACME:2018yjb} & (Advanced ACME~\cite{Advanced-ACME}) \\
\hline
\end{tabular}
}
\caption{Experimental results for $a_\mu$ and upper limits for $\mathcal B (\mu \to e \gamma)$, $\vert d_\mu/e \vert$ and $\vert d_e/e \vert$.
}
\label{observables}
\end{table}

We lay out the paper as follows.
In the next section, we will review the minimal particle content in G2HDM and write down the relevant interactions required for 
the one-loop computation for the form factors of the radiative decays $l_i \to l_j \gamma$.
In Section 3, we compute the magnetic and electric dipole form factors for the radiative decays. 
In the case of $i=j$ we also obtain the anomalous magnetic dipole moment and electric dipole moment for the lepton $l_i$. 
We will show that the electric dipole moment of the lepton vanishes identically at one-loop in G2HDM.
Numerical analysis for $\mu \to e \gamma$ and $\Delta a_\mu$ is presented in Section 4. 
We also present the impact of the viable parameter space on the spin-independent cross section for the sub-GeV dark matter direct search experiments.
We conclude in Section 5. 

Analytical formulas for the form factors and the associated loop integrals are given in Appendix A.
In Appendix B, we show that the form factors of the SM $W$ boson loop 
obtained in the unitary and 't Hooft-Feynman gauges are equivalent. Discrepancies between our results and existing ones in the 
literature are clarified in the Appendices. Some relevant Feynman rules in G2HDM are shown in Appendix C. 
In Appendix D, we demonstrate the well-known fact that only the magnetic and electric dipole moment form factors are relevant for the computations 
of the on-shell amplitude of $l_i \to l_j \gamma$.



\section{Minimal G2HDM}
\label{G2HDM}

In this section, we will briefly review the minimal G2HDM studied recently in~\cite{Ramos:2021omo,Ramos:2021txu,Tran:2022yrh}. 
The original model based on augmenting the SM electroweak gauge group $SU(2)_L \times U(1)_Y$ by a hidden gauge sector $SU(2)_H \times U(1)_X$
was introduced in Ref.~\cite{Huang:2015wts}.
The main idea of G2HDM is to group the two Higgs doublets $H_1$ and $H_2$ in inert 2HDM (I2HDM) together to form a 
bifundamental irreducible representation of $SU(2)_L \times SU(2)_H$.
Various refinements~\cite{Arhrib:2018sbz,Huang:2019obt,Chen:2019pnt} and 
collider phenomenology~\cite{Chen:2018wjl,Huang:2017bto,Huang:2015rkj} were pursued subsequently with the same particle content as 
the original model where the DM candidate is a complex scalar $\mathcal D$. 
In this work, as in~\cite{Ramos:2021omo,Ramos:2021txu,Tran:2022yrh},
we will drop the triplet field $\Delta_H$ of the extra $SU(2)_H$ in the original model 
and propose the complex gauge boson field $\mathcal W^{\prime (p,m)}$ as DM candidate rather than the complex scalar $\mathcal D$.
For convenience, the scalar and fermion contents and their quantum numbers as well as $h$-parity in the model  are tabulated in Table~\ref{tab:quantumnosscalar} and ~\ref{tab:quantumnosfermion} respectively.
Our convention for the electric charge $Q$ (in unit of $e$) is
$Q = T^3_L + Y$ where $T^3_L$ is the third component of the $SU(2)_L$ generators  
and  $Y$ is the hypercharge.
$S$ is the scalar field introduced to implement the Stueckelberg mechanism to provide a mass 
for the $U(1)_X$ gauge boson~\cite{Kors:2004dx,Kors:2005uz,Feldman:2006wb,Feldman:2007wj}.

\begin{table}[htbp!]
\begin{tabular}{|c|c|c|c|c|c|c|}
\hline
Scalar & $SU(2)_L$ & $SU(2)_H$ & $U(1)_Y$ & $U(1)_X$ & $h$-parity\\
\hline\hline
$H=\left( H_1 \;\; H_2 \right)^{\rm T}$ & 2 & 2 & $\frac{1}{2}$ & $\frac{1}{2}$ & $(+,-)$ \\
$\Phi_H=\left( \Phi_1 \;\; \Phi_2 \right)^{\rm T}$ & 1 & 2 & 0 & $\frac{1}{2}$ & $(-,+)$\\
$S$ & 1 & 1 & 0 & 0 & $+$ \\
\hline
\end{tabular}
\caption{Higgs scalars in the minimal G2HDM and their quantum number assignments.
}
\label{tab:quantumnosscalar}
\end{table}

\begin{table}[htbp!]
\begin{tabular}{|c|c|c|c|c|c|c|}
\hline
Fermion &  $SU(3)_C$ & $SU(2)_L$ & $SU(2)_H$ & $U(1)_Y$ & $U(1)_X$ & $h$-parity \\
\hline \hline
$Q_L=\left( u_L \;\; d_L \right)^{\rm T}$ & 3 & 2 & 1 & $\frac{1}{6}$ & 0 & $(+,+)$ \\
$U_R=\left( u_R \;\; u^H_R \right)^{\rm T}$  & 3 & 1 & 2 & $\frac{2}{3}$ & $\frac{1}{2}$ & $(+,-)$\\
$D_R=\left( d^H_R \;\; d_R \right)^{\rm T}$ & 3 & 1 & 2 & $-\frac{1}{3}$ & $-\frac{1}{2}$ & $(-,+)$ \\
\hline
$u_L^H$ & 3 & 1 & 1 & $\frac{2}{3}$ & 0 & $-$ \\
$d_L^H$ & 3 & 1 & 1 & $-\frac{1}{3}$ & 0 & $-$ \\
\hline
$L_L=\left( \nu_L \;\; e_L \right)^{\rm T}$ & 1 & 2 & 1 & $-\frac{1}{2}$ & 0 & $(+,+)$\\
$N_R=\left( \nu_R \;\; \nu^H_R \right)^{\rm T}$ & 1 & 1 & 2 & 0 & $\frac{1}{2}$ & $(+,-)$\\
$E_R=\left( e^H_R \;\; e_R \right)^{\rm T}$ & 1 & 1 & 2 &  $-1$  &  $-\frac{1}{2}$ & $(-,+)$\\
\hline
$\nu_L^H$ & 1 & 1 & 1 & 0 & 0 & $-$ \\
$e_L^H$ & 1 & 1 & 1 & $-1$ & 0 & $-$\\
\hline
\end{tabular}
\caption{Fermions in the minimal G2HDM and their quantum number assignments.}
\label{tab:quantumnosfermion}
\end{table}

\subsection{Higgs Potential and Spontaneous Symmetry Breaking}
\label{HiggsPotential}

The most general Higgs potential which is invariant under both $SU(2)_L\times U(1)_Y \times SU(2)_H \times  U(1)_X$  
can be written down as follows
\begin{align}\label{eq:V}
V = {}& - \mu^2_{\Phi}   \Phi_H^\dag \Phi_H  + \la_\Phi \lee \Phi_H^\dag \Phi_H  \rii^2
 - \mu^2_H   \left(H^{\alpha i}  H_{\alpha i} \right) 
+  \lambda_H \left(H^{\alpha i}  H_{\alpha i} \right)^2  \nonumber  \\
{}& + \frac{1}{2} \lambda'_H \epsilon_{\alpha \beta} \epsilon^{\gamma \delta} 
 \left(H^{ \alpha i}  H_{\gamma  i} \right)  \left(H^{ \beta j}  H_{\delta j} \right)
+\lambda_{H\Phi} \lee H^\dag H  \rii  \lee \Phi_H^\dag \Phi_H \rii  \\
{}&+ \lambda^\prime_{H\Phi} \lee H^\dag \Phi_H  \rii  \lee \Phi_H^\dag H \rii, \nonumber
\end{align}
where  ($\alpha$, $\beta$, $\gamma$, $\delta$) and ($i$, $j$) refer to the $SU(2)_H$ and $SU(2)_L$ indices respectively, 
all of which run from one to two, and $H^{\alpha i} = H^*_{\alpha i}$.

To study spontaneous symmetry breaking (SSB) in the model, we parameterize the Higgs fields according to standard practice
\begin{eqnarray}
\label{eq:scalarfields}
H_1 = 
\begin{pmatrix}
G^+ \\ \frac{ v + h_{\rm SM}}{\sqrt 2} + i \frac{G^0}{\sqrt 2}
\end{pmatrix}
, \,
H_2 = 
\begin{pmatrix}
\mathcal H^+  \\ \mathcal H_2^0 
\end{pmatrix}
, \,
\Phi_H = 
\begin{pmatrix}
G_H^p  \\ \frac{ v_\Phi + \phi_H}{\sqrt 2} + i \frac{G_H^0}{\sqrt 2}
\end{pmatrix}
\; \; \;
\end{eqnarray}
where $v$ and $v_\Phi$ are the only non-vanishing vacuum expectation values (VEVs)
in $H_1$ and $\Phi_{H}$ fields respectively. $H_2$ does not develop VEV as in the case of I2HDM.

Theoretical constraints like bounded from below and perturbative unitarity of the above scalar potential can be found in our previous works~\cite{Ramos:2021omo,Ramos:2021txu}.

\subsection{Interaction Lagrangian}

Besides the unitary Pontecorvo-Maki-Nakagawa-Sakata (PMNS) neutrino mixing matrix
\beq
V_{\rm PMNS} \equiv \left( U_\nu^L \right)^\dagger U^L_l \; ,
\eeq
defined in the left-handed lepton sector, we also need to introduce the following unitary mixing matrices in the right-handed lepton sector in G2HDM,
\begin{align}
V^H_l & \equiv \left( U^R_l \right)^\dagger U^R_{l^H} \, , \nonumber \\
V_\nu^{H } & \equiv \left( U^R_\nu \right)^\dagger U^R_{\nu^H} \; .
\end{align}

\begin{figure}[t] 
   \centering
   \includegraphics[scale = 0.43]{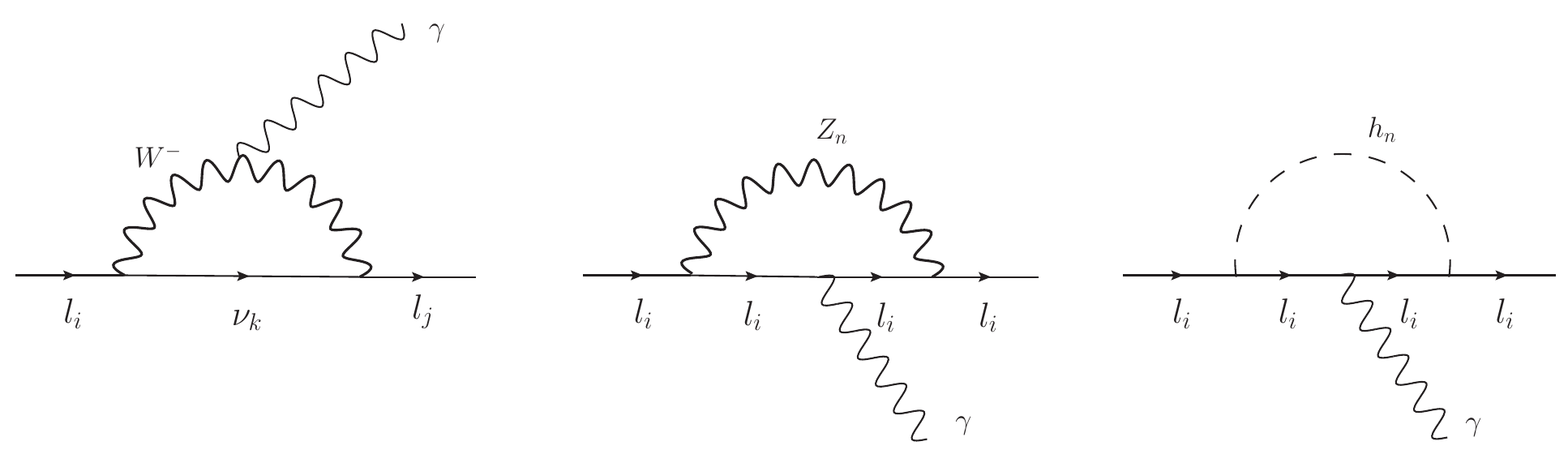}
   \caption{ The one-loop SM-like contribution to $l_i - l_j - \gamma$ vertex from the SM $W$ boson diagram (left panel), and contributions to $l_i - l_i - \gamma$ vertex from $\left\{ Z_n \right\}$ diagram (center panel) and $\left\{ h_n \right\}$ diagram (right panel) in G2HDM.} 
   \label{fig:li-lj-gamma-SM-like}
\end{figure}

\begin{figure}[t] 
   \centering
 \includegraphics[scale=0.43]{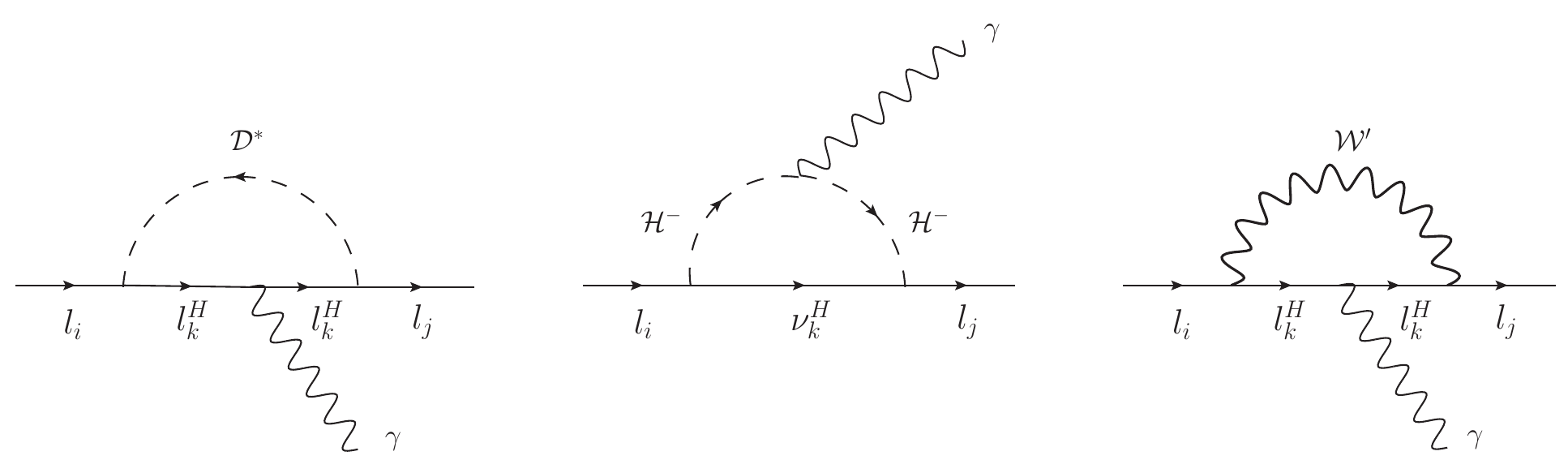}  
   \caption{Three new contributions of $\mathcal D$, $\mathcal H^+$ and $\mathcal W'$ to $l_i - l_j - \gamma$ vertex in G2HDM.}
   \label{fig:li-lj-gamma-G2HDM}
\end{figure}

There are altogether 6 one-loop contributions to the $l_i - l_j  - \gamma$ vertex in the minimal  G2HDM.
The Feynman diagrams are shown in Figs.~(\ref{fig:li-lj-gamma-SM-like}) and~(\ref{fig:li-lj-gamma-G2HDM}).
Here the self-energy diagrams are not explicitly shown. 
However they contribute to the $\gamma^\mu$ and $\gamma^\mu \gamma_5$ form factors in the amplitude which are important for the cancellation of ultraviolet divergences and the maintenance of gauge invariance.
Fig.~(\ref{fig:li-lj-gamma-SM-like}) is the SM-like contributions with all $h$-parity even particles circulating inside the loop, 
while Fig.~(\ref{fig:li-lj-gamma-G2HDM}) is the new contributions from G2HDM with all $h$-parity odd particles circulating inside the loop.
The QED vertex for a photon couples with $W^\pm$, $l_i$, $l_i^H$ and $\mathcal H^\pm$ are standard, they can be found in many textbooks
and will be omitted in what follows.

The first diagram in Fig.~(\ref{fig:li-lj-gamma-SM-like}) is the contribution from the SM charged $W^\pm$ boson. The relevant interaction Lagrangian is
\beq
\label{LSMW}
\mathcal L^W \supset \frac{g}{2 \sqrt 2} \sum_{i,k}\left( V_{\rm PMNS} \right)_{ki}  \bar \nu_k \gamma^\mu \left( 1 - \gamma_5 \right) l_i  W^+_\mu + {\rm H. c.} \; .
\eeq

The second diagram in Fig.~(\ref{fig:li-lj-gamma-SM-like}) is the contribution from the neutral gauge bosons $\{ Z_n \}$. 
The relevant interaction Lagrangian is 
\beq
\label{LZn}
\mathcal L^{\{ Z_n \}} \supset \sum_n \sum_{i} \bar l_i  \gamma_\mu \left( C_{V  n} + C_{A n} \gamma_5 \right)  l_i Z^\mu_n \; ,
\eeq
where $C_{V n}$ and $C_{A n}$ are the vector and axial-vector coupling constants. Based on lepton universality, these couplings are independent of the charged lepton flavor $i$. 
Their expressions are given by $C_{Vn}=(C_{Ln}+C_{Rn})/2$ and $C_{An} = (-C_{Ln}+C_{Rn})/2$ with 
\bea
C_{Ln} & = & \frac{g}{\cos \theta_{\rm W}} \left(  - \frac{1}{2} + \sin^2\theta_{\rm W} \right) O^N_{1n} \;, \\
C_{Rn} & = & \frac{g}{\cos \theta_{\rm W}} \sin^2\theta_{\rm W} O^N_{1n} - \frac{1}{2} g_H O^N_{2n} - \frac{1}{2} g_X O^N_{3n} \; ,
\eea
where $\theta_{\rm W}$ is the weak mixing angle, $g_X$ and $g_H$ are gauge couplings of the $U(1)_X$ and $SU(2)_H$, respectively.
$O^N$ is a $3\times 3$ orthogonal matrix that diagonalizes the following mass matrix in the basis of $(Z^{\rm SM}, \mathcal W'_3, X)$
\be
\mathcal M_Z^2 =  
\begin{pmatrix}
m^2_{Z} & - \frac{1}{2} g_H v m_{Z} & - { \frac{1}{2} } g_X v m_{Z} \\
 - \frac{1}{2} g_H v m_{Z} & m^2_{\mathcal W^\prime} & { \frac{1}{4} } g_H g_X v_{-}^2 \\
- {\frac{1}{2} } g_X v m_{Z} &   { \frac{1 }{4} } g_H g_X v_{-}^2 & {\frac{1}{4} } g_X^2 v_{+}^2 + M_X^2
\end{pmatrix} \; ,
\label{MsqZs}
\ee
where 
\bea
 \label{mzsm}
 m_{Z} & = &  \frac{1 }{2}  v \sqrt{ g^2 + g^{\prime \, 2} }\; , \\
 \label{mwprime}
 m_{\mathcal W^\prime} & = & \frac{1}{2} g_H \sqrt{ v^2 + v_\Phi^2 } \; ,\\
 \label{vsqplusminus}
 v_{\pm}^2 & = & \left( v^2 \pm v^2_\Phi \right) \; ,
\eea
and $M_X$ is the Stueckelberg mass for the $U(1)_X$. 
We denote the physical mass eigenstates as $Z_n \, (n=1,2,3)$ with the mass ordering $M_{Z_1} \geq M_{Z_2} \geq M_{Z_3}$. 
In the parameter space choice in our numerical work, $Z_1$ will be identified as the $Z$ boson of 91.1876 GeV~\cite{Zyla:2020zbs} observed at LEP,
$Z_2$ is the dark $Z'$ and $Z_3$ is the dark photon $\gamma'$ (or $A'$ in some literature). They all have even $h$-parity.

The third diagram in Fig.~(\ref{fig:li-lj-gamma-SM-like}) is the contribution from the neutral Higgs bosons $\{ h_n \}$. 
The relevant interaction Lagrangian is 
\beq
\label{Lhn}
\mathcal L^{\{ h_n \}} \supset - \sum_n \sum_{i} \left( O^H \right)_{1 n} \frac{m_i}{v} \bar l_i  l_i h_n \; ,
\eeq
where $O^H$ is the mixing matrix between $h_{\rm SM}$ and $\phi_H$,
\be
\left(
\begin{matrix}
h_{\rm SM} \\
\phi_H
\end{matrix}
\right)
= 
O^H \cdot 
\left(
\begin{matrix}
h_1 \\
h_2
\end{matrix}
\right)
=
\left( 
\begin{matrix}
 \cos \theta_1  &  \sin \theta_1 \\
- \sin \theta_1  &  \cos \theta_1 
\end{matrix}
\right) \cdot 
\left(
\begin{matrix}
h_1 \\
h_2
\end{matrix}
\right)
\; .
\ee
The mixing angle $\theta_1$ is given by
\be
\tan 2 \theta_1 = \frac{\lambda_{H\Phi} v v_\Phi }{ \lambda_\Phi v^2_\Phi - \lambda_H v^2 } \; .
\ee
The masses of $h_1$ and $h_2$ are given by
\bea
m_{h_1,h_2}^2 &=& \lambda_H v^2 + \lambda_\Phi v_\Phi^2 
\mp \sqrt{\lambda_H^2 v^4 + \lambda_\Phi^2 v_\Phi^4 + \left( \lambda^2_{H\Phi}  - 2 \lambda_H \lambda_\Phi \right) v^2 v_\Phi^2 } \; .
\eea
Depending on its mass, either $h_1$ or $h_2$ is identified as the observed Higgs boson $h$ at the Large Hadron Collider (LHC).
Currently the most precise measurement of the Higgs boson mass is $m_{h} = 125.38 \pm 0.14$ GeV~\cite{CMS:2020xrn}.
In this work, we will identify the lighter state $h_1$ as $h$.

Since the gauge and Yukawa couplings in (\ref{LZn}) and (\ref{Lhn}) respectively are all real and flavor diagonal, 
there are no contributions to $l_i \to l_j \gamma \, (i \neq j)$ and electric dipole moment of $l_i$
from the interactions $\mathcal L^{\{ Z_n \}}$ and $\mathcal L^{\{ h_n \}}$.
The only non-vanishing contribution to $l_i \to l_j \gamma \, (i \neq j)$ in SM at one-loop is the charged $W^\pm$ from $\mathcal L^W$ in (\ref{LSMW}).
However it is well known that its amplitude is suppressed by the squared of neutrino masses due to GIM-like mechanism in the lepton sector.
Furthermore, due to the unitarity of $V_{\rm PMNS}$, $d_{l_i}(W^\pm)$ also vanishes at one-loop. See Appendix A.

Next we turn to the new contributions in G2HDM.

The first diagram in Fig.~(\ref{fig:li-lj-gamma-G2HDM}) is the contribution from the dark Higgs $\mathcal D$ which is a linear combination of two odd 
$h$-parity components $\mathcal H_2^0$ and $G_H^m$,
\beq
\mathcal D = \cos\theta_2 \mathcal H_2^{0} + \sin\theta_2 G_H^m \; ,
\eeq
where $\theta_2$ is a mixing angle giving by
\beq
\label{theta2def}
\tan 2 \theta_2 = \frac{2 v v_\Phi}{v_\Phi^2 - v^2} \; .
\eeq
The mass of $\mathcal D$ is
\be
m_{\mathcal D}^2 = \frac{1}{2} \lambda^\prime_{H\Phi} v_+^2\; ,
\ee
where $v^2_+$ is defined in (\ref{vsqplusminus}).

The relevant interaction Lagrangian is given by
\beq
{\mathcal L}^{\mathcal D} \supset \sum_{i,j} \overline{ l^H_i } \left( y^{\mathcal D}_{S \, ij}   +
y^{\mathcal D}_{P \, ij} \gamma_5  \right) l_j \mathcal D^* + {\rm H.c.} \; ,
\eeq
where the scalar and pseudoscalar Yukawa couplings $y^{\mathcal D}_{S\, ij}$ and $y^{\mathcal D}_{P \, ij}$ are given by
\beq
\label{YukD}
y^{\mathcal D}_{S/P \, ij} = \pm \frac{\sqrt 2}{2 v} \cos\theta_2 \left( V_l^{H\, \dagger} M_l \right)_{ij}
+ \frac{\sqrt 2}{2 v_\Phi} \sin\theta_2 \left( M_{l^H} V_l^{H\, \dagger} \right)_{ij} \; ,
\eeq
with $M_l = {\rm diag} \left( m_e, m_\mu, m_\tau \right)$ and
$M_{l^H} = {\rm diag} \left( m_{l^H_1}, m_{l^H_2}, m_{l^H_3} \right)$. 
Note that the ordering of the mass matrices are important in the 
Yukawa couplings  (\ref{YukD}).
From  (\ref{YukD}), one obtains 
\be
\label{yPsyS}
y^{\mathcal D \, *}_{P \, ki} y^{\mathcal D}_{S \, ki} = \frac{1}{2}\left| \left( V_l^{H} \right)_{ik} \right|^2  \left(\frac{ m_{l^H_k}^2 } {v_\Phi^2}\sin^2 \theta_2  - \frac{ m_{l_i}^2} {v^2}  \cos^2 \theta_2 \right) \, .
\ee
Thus ${\rm Im} \left( y^{\mathcal D \, *}_{P \, ki} y^{\mathcal D}_{S \, ki}  \right) = 0$.
We don't expect the complex Yukawa couplings in 
$\mathcal L^{\mathcal D}$ to give rise a non-vanishing electric dipole moment $d_{l_i}$ at one-loop, as shown in Appendix A.

The second diagram in Fig.~(\ref{fig:li-lj-gamma-G2HDM}) is the contribution from the dark charged Higgs $\mathcal H^\pm$ which has odd $h$-parity and a mass given by
\be
m^2_{\mathcal H^\pm} = \frac{1}{2} \left( \lambda^\prime_{H\Phi} v^2_\Phi - \lambda^\prime_H v^2 \right) \; .
\ee
The relevant interaction Lagrangian is given by
\beq
{\mathcal L}^{\mathcal H} \supset \sum_{i,j} \overline{ \nu^H_i } \left(  y^{\mathcal H}_{S \, ij}    +
y^{\mathcal H}_{P \, ij}  \gamma_5 \right)  l_j  \mathcal H^+ + {\rm H.c.} \; ,
\eeq
where the scalar and pseudoscalar Yukawa couplings $y^{\mathcal H}_{S \, ij}$ and $y^{\mathcal H}_{P \, ij}$  are given by 
\beq
\label{YukawaH}
y^{\mathcal H}_{S/P \, ij}  = \pm \frac{\sqrt 2}{2  v} \left( V_\nu^{H \, \dagger} M_\nu V_{\rm PMNS} \right)_{ij} \; ,
\eeq
with $M_\nu = {\rm diag}\left( m_{\nu_1}, m_{\nu_2}, m_{\nu_3} \right)$. Since the Yukawa couplings $y^{\mathcal H}_{S \, ij} $ and $y^{\mathcal H}_{P \, ij}$ 
are related, we expect $d_{l_i}(\mathcal H^\pm) = 0$ at one-loop. See Appendix A for detail.

The third diagram in Fig.~(\ref{fig:li-lj-gamma-G2HDM}) is the contribution from the vector dark matter $\mathcal W^{\prime (p,m)}$ 
$(\equiv (\mathcal W'_1 \mp i \mathcal W'_2)/\sqrt 2)$
which is assumed to be the lightest $h$-parity odd particle in G2HDM. 
Its mass is given by (\ref{mwprime}).
The relevant interaction Lagrangian is given by
\beq
{\mathcal L}^{\mathcal W'} \supset \sum_{i,j}   \overline{ l^H_i } \gamma^\mu \left( g^{\mathcal W'}_{V \, ij}  +
g^{\mathcal W'}_{A \, ij}  \, \gamma_5  \right) l_j \mathcal W^{\prime \, p}_\mu  + {\rm H.c.} \; ,
\eeq
where the vector and axial gauge couplings $g^{\mathcal W'}_{V \, ij}$ and $g^{\mathcal W'}_{A \, ij}$ are given by
\beq
\label{gaugecouplingWp}
g^{\mathcal W'}_{V \, ij} = g^{\mathcal W'}_{A \, ij} = \frac{g_H}{2 \sqrt 2} \left( V^H_l \right)^\dagger_{ij} 
\; .
\eeq
Since the vector and axial vector couplings $g^{\mathcal W'}_{V \, ij} $ and $g^{\mathcal W'}_{A \, ij}$ are the same, we expect $d_{l_i}(\mathcal W') = 0$ at one-loop.
(See Appendix A.)
This is analogous to the SM charged $W^\pm$ case where the vector and axial vector couplings are opposite sign to each other, there 
as is well-known we have $d_{l_i}(W^\pm) = 0$ at one-loop too.

In summary, we expect all the new flavor non-diagonal complex couplings from $\mathcal L^{\mathcal D}$, $\mathcal L^{\mathcal H}$ and $\mathcal L^{\mathcal W'}$ 
in G2HDM can give rise to contributions to $l_i \to l_j \gamma \, (i \neq j)$. 
Certainly they will all give non-vanishing contributions to $a_{l_i}$ but not $d_{l_i}$ at one-loop. 
The relevant Feynman rules are given in Appendix C.


\begin{figure}[htbp] 
   \centering
   \includegraphics[scale=0.5]{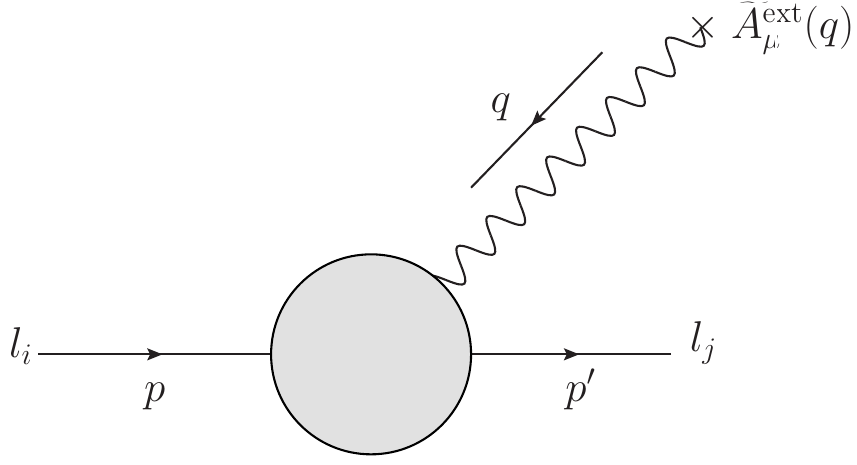} 
   \caption{Feynman Diagram for the $l_i-l_j-\gamma$ vertex.}
   \label{fig:Feynman-vertex}
\end{figure}

\section{Magnetic and Electric Dipole Form Factors}

The Lorentz invariant amplitude for a charged lepton $l_i$ of flavor $i$ scatters with an electromagnetic background field $\widetilde{\mathcal A}^{\rm ext}_\mu (q)$ 
to become another charged lepton $l_j$ of flavor $j$ as depicted in Fig.~(\ref{fig:Feynman-vertex}) is given by 
\footnote{In general, the amplitude has six Lorentz decomposition form factors, however, due to the gauge invariance, 
only $\sigma^{\mu\nu} q_{\nu}$ and $\sigma^{\mu\nu} q_{\nu} \gamma_5$ terms are retained for on-shell photon where $q^2 = 0$ (see Appendix D for a detailed discussion).}
\beq
\label{matrixelement}
i \mathcal M_{ji} = \overline{ u_j }\left( p' \right) \left( - i e \Gamma_{ji}^\mu \right) u_i \left( p \right) \widetilde{\mathcal A}^{\rm ext}_\mu \left(  q \right)  \; ,
\eeq
with  $-e \, (e > 0)$ and $m_i$ are the electric charge and mass of $l_i$ respectively, $q = (p' - p)$ is the momentum transfer, 
and the vertex function $\Gamma^\mu_{ji}$ can be decomposed as
\beq
\label{vertex}
\Gamma_{ji}^\mu =   i \sigma^{\mu\nu} q_\nu \frac{m_i}{2} \left( A^M_{ji} + i A^E_{ji} \gamma_5 \right) \; ,
\eeq
with $A^M_{ji}$ and $A^E_{ji}$ related to the transition magnetic and electric dipole form factors 
respectively~\footnote{For ease of comparisons of their analytical expressions presented in the Appendices, 
we use the same notations $A^{M}_{ji}$ and $A^E_{ji}$ as in~\cite{Lindner:2016bgg}. And they are understood to be evaluated at $q^2=0$.}.

The effective Lagrangian that can reproduce the matrix element
(\ref{matrixelement}) with the associated vertex (\ref{vertex}) is 
\beq
\label{Leff}
\mathcal L_{\rm eff}   =  - \frac{1 }{4} e \, m_i  \overline{ l_j}  \sigma^{\mu \nu} \left( A^M_{ji} + i \gamma_5 A^E_{ji} \right) l_i F^{\rm ext}_{\mu\nu}\; ,
\eeq
where $F^{\rm ext}_{\mu\nu}$ is the electromagnetic background field strength.

The above form factors $A^M_{ji}$ and $A^E_{ji}$ enable us to compute the decay rate for the process $l_i \to l_j \gamma \; (i \neq j)$
with the following spin-averaged matrix element squared
\beq
\overline{ \sum } \vert  \mathcal M_{ji} \vert^2 = 
\frac{e^2}{2} m_{i}^6 \left( 1 - \frac{m_{j}^2}{m_{i}^2} \right)^2 \left( \vert A^M_{ji} \vert^2 + \vert A^E_{ji} \vert^2  \right) \; .
\eeq
We thus obtain the decay rate and branching ratio for $l_i \to l_j \gamma$ 
\begin{align}
\Gamma \left( l_i \to l_j \gamma \right) & = \frac{1}{32 \pi} m^5_{i} \left( 1 - \frac{ m^2_{j} }{ m^2_{i} } \right)^3 
e^2 \left( \vert A^M_{ji} \vert^2 + \vert A^E_{ji} \vert^2 \right)  \; ,\\
\mathcal B \left( l_i \to l_j \gamma \right) & = \frac{ \Gamma \left( l_i \to l_j \gamma \right) } { \Gamma \left( l_i \to l_j \nu_i \overline{\nu_j} \right) } \cdot
\frac { \Gamma \left( l_i \to l_j \nu_i \overline{\nu_j} \right) } {\Gamma_{l_i} } \; ,
\end{align}
where~\footnote{See for example the Appendix in the textbook {\it Collider Physics}, Updated Edition, CRC Press 1996, by Barger and Phillips.}
\beq
\Gamma \left( l_i \to l_j \nu_i \overline{\nu_j} \right)   = \frac{G_F^2 m_{i}^5}{192 \pi^3}  f\left( \frac{m_j}{m_i} \right) \;, \\
\eeq
with $G_F$ is the Fermi constant and
\beq
 f\left( x \right)  = \left(
1 - 8 \, x^2 + 8 \, x^6  - x^8 - 24 \, x^4 \log  x  \right) \; .
\eeq
Therefore
\beq
\mathcal B \left( l_i \to l_j \gamma \right) = \frac{3 \left( 4 \pi \right)^3 \alpha_{\rm EM }}{ 2 G_F^2} \cdot 
\frac{ \left( 1 - \frac{ m^2_{j} }{ m^2_{i} } \right)^3 }{f\left( \frac{m_j}{m_i}\right)} \cdot
\left( \vert A^M_{ji} \vert^2 + \vert A^E_{ji} \vert^2\right)  \cdot
\mathcal B \left( l_i \to l_j \nu_i \overline{\nu_j} \right)  \;,
\eeq
where $\alpha_{\rm EM} = e^2/(4\pi)$.
For $\mu \to e \overline{\nu_e} \nu_\mu $,  
\beq
\mathcal B \left( \mu \to e \overline{\nu_e} \nu_\mu  \right) \approx 100 \% \; .
\eeq

For $l_i \to l_j \gamma$ ($i\neq j$) in G2HDM there are 4 distinct non-vanishing contributions to each $A^M$ and $A^E$,
\beq
A^{M/E}_{ji}  = A^{M/E}_{ji} \left( W \right) + A^{M/E}_{ji} \left( \mathcal D \right) + A^{M/E}_{ji} \left( \mathcal H \right) + A^{M/E}_{ji} \left( \mathcal W' \right) \; .
\eeq
As is well known the SM contribution $A^{M/E}_{ji} \left( W \right)$ for $i\neq j$ 
from the $W$ boson loop is vanishingly small and many orders below the current experimental sensitivities.

The anomalous magnetic dipole moment $a_{l_i}$ of the charged lepton $l_i$ can be identified as the coefficient of $\left( e/2\, m_{i} \right) i \sigma^{\mu\nu}q_\nu$ in the vertex $e\Gamma^\mu_{ii}$ of (\ref{vertex}), {\it i.e.}
\be
a_{l_i}  = m_i^2 A^M_{ii} \; , \;\;\;\;\;\; ({\rm no \; sum \; on \;} i) \; .
\ee

The electric dipole moment $d_{i_i}$ of the charged lepton $l_i$ is given by
\be
\label{EDMdef}
\frac{d_{l^i}}{e}   =   \frac{m_{i}}{2} A^E_{ii} \; , \;\;\;\;\;\; ({\rm no \; sum \; on \;} i) \; .
\ee

For the anomalous magnetic dipole moment of $l_i$ in G2HDM, 
besides the well-known QED contribution $a_{l_i}(\gamma) = \alpha_{\rm EM}/2\pi$, 
there are in general 6 distinct electroweak contributions to the $a_{l_i}$,
\begin{align}
a_{l_i} & = \frac{\alpha_{\rm EM}}{2\pi} +m_i^2 A^M_{ii} \; ,\;\; ({\rm no \; sum \; on} \; i) \; , \nonumber \\
A^{M}_{ii} & = A^{M}_{ii} \left( W \right) + A^{M}_{ii} \left( \left\{ Z_n \right\} \right)  + A^{M}_{ii} \left( \left\{ h_n \right\}  \right)  
+ A^{M}_{ii} \left( \mathcal D \right) + A^{M}_{ii} \left( \mathcal H \right) + A^{M}_{ii} \left( \mathcal W' \right) \; .
\end{align}

Analytical one-loop expressions for $A^{M/E}_{ji}$ are given in Appendix A. There one will see all the $A^E_{ii}$s vanish at one-loop in G2HDM, hence the electric dipole moment $d_{l_i}$ of $l_i$ vanish too 
according to (\ref{EDMdef}).
These form factors $A^{M/E}_{ji}$ were also computed for general couplings in~\cite{Lindner:2016bgg}.  
Aside from an overall factor of 2 in the form factors, we will discuss some minor discrepancies in the loop integrals between our results and~\cite{Lindner:2016bgg} in Appendix A.

\section{Numerical results}
In this section, we show numerical results for the cLFV process $\mu \to e \gamma$ 
and muon anomalous magnetic dipole moment 
with the parameter space in the model chosen 
to satisfy the current constraints for a sub-GeV non-abelian vector DM $\mathcal W'$. 
In particular, the scan data are adapted from Ref.~\cite{Tran:2022yrh} in which the theoretical constraints on the scalar potential~\cite{Ramos:2021omo,Ramos:2021txu}, 
signal strength measurements from the LHC~\cite{Sirunyan:2018koj,Aad:2019mbh,ATLAS:2021vrm}, dark photon physics~\cite{ATLAS:2019erb,Fabbrichesi:2020wbt}, 
electroweak precision measurements~\cite{Zyla:2020zbs} including the recent $W$ boson mass measurement at the CDF II~\cite{CDF:2022hxs} and 
constraints from DM searches including the DM relic density measured from Planck collaboration~\cite{Aghanim:2018eyx}, 
DM direct detections~\cite{Angloher:2017sxg,Agnes:2018ves,Aprile:2019xxb} 
and Higgs invisible decays constraint from the LHC~\cite{ATLAS:2022yvh}.
For the data points that satisfy the above mentioned constraints, the total DM annihilation cross section is of order $10^{-32} \, {\rm cm^3}\cdot s^{-1}$ or below that is much lower than the current DM indirect detection constraints.

We set the mixing matrices in the right-handed lepton sector in the model to be 
\be
V_{l}^H = V_{\nu}^H = V_{\rm PMNS},
\ee
where $V_{\rm PMNS}$ is parameterized as 
\be
\label{pmnsmatrix}
V_{\rm PMNS} \equiv
\begin{pmatrix}
c_{12} c_{13} & s_{12} c_{13} & s_{13} e^{- i\delta_{\rm CP}}\\
-s_{12} c_{23} - c_{12} s_{23} s_{13} e^{ i\delta_{\rm CP}} & c_{12} c_{23} - s_{12} s_{23} s_{13}  e^{ i\delta_{\rm CP}}  & s_{23} c_{13} \\ 
s_{12} s_{23} - c_{12} c_{23} s_{13} e^{ i\delta_{\rm CP}} & -c_{12} s_{23} - s_{12} c_{23} s_{13}  e^{ i\delta_{\rm CP}}  & c_{23} c_{13} 
\end{pmatrix},
\ee
where $s_{ij}$ and $c_{ij}$ stand for $\sin \theta_{ij} $ and $\cos \theta_{ij} $ respectively, and $\delta_{\rm CP}$ is a Dirac CP violating phase. 
The current best-fit values using a normal ordering are given by \cite{Esteban:2020cvm}:
$\theta_{12} = {33.44^{\circ}}^{ + 0.77^{\circ}} _{-0.74^{\circ}}$, 
$\theta_{23} = {49.2^{\circ}}^{ + 1.0^{\circ}} _{-1.3^{\circ}}$, 
$\theta_{13} = {8.57^{\circ}}^{ + 0.13^{\circ}} _{-0.12^{\circ}}$, and 
 $\delta_{\rm CP} = {194^{\circ}}^{ + 52^{\circ}} _{-25^{\circ}}$.
 We also set the heavy hidden lepton masses to be 
 \be
 M_{\nu^H}  = M_{l^H} = {\rm diag} \left( m_{l^H}, m_{l^H} + \Delta m_{l^H},  m_{l^H} + \Delta m_{l^H} \right)
 \ee
 where the second and third generations are assumed degenerate, and $\Delta m_{l^H}$ is a mass splitting between the first and second (third) generations. 
 
 \begin{figure}[tb]
         \centering
	\includegraphics[width=0.47\textwidth]{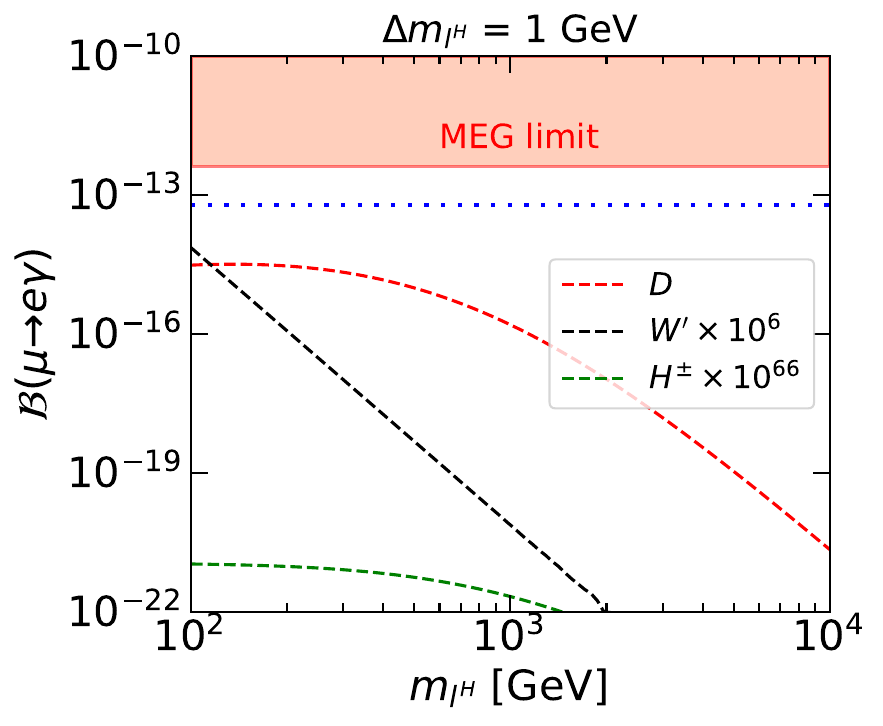}
	\includegraphics[width=0.47\textwidth]{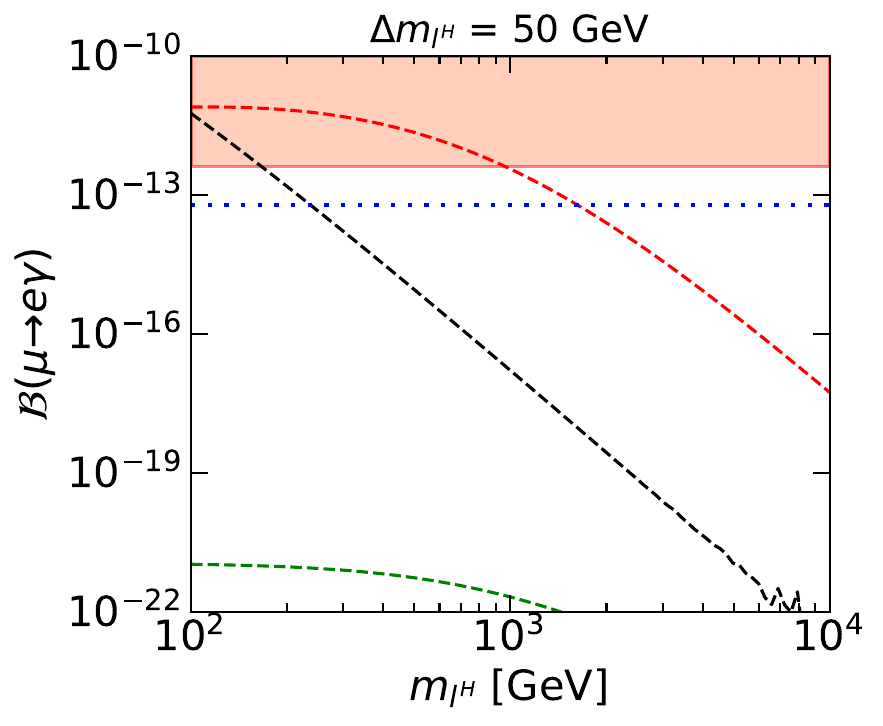}
	\includegraphics[width=0.47\textwidth]{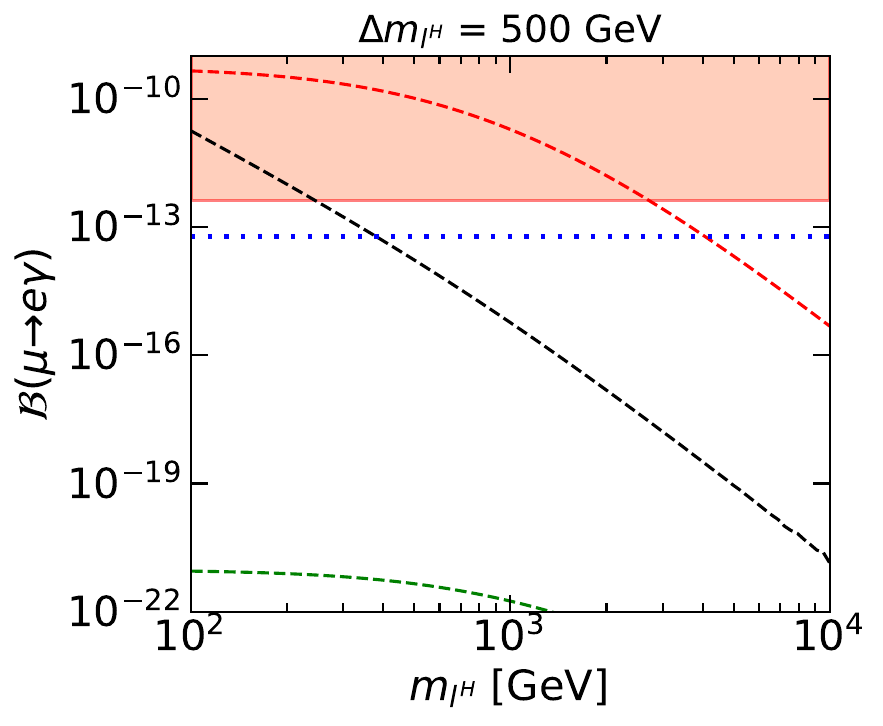}
	\caption{ \label{fig:brmu2e-mlH} Branching ratio of $\mu \to e \gamma$ as a function of the heavy hidden lepton mass $m_{l^H}$. 
	Other parameters in the model are set to be 
	$m_{h_2} = 292.50$ GeV, $m_{\mathcal D} = 766.07$ GeV, 
	$m_{\mathcal H^{\pm}} = 848.13$ GeV, 
	$m_{\mathcal W'} = 1.0$ GeV, $\theta_{1} = 0.030$ rad, 
	$\theta_{2} = 0.056$ rad, 
	$M_{X} = 1.96$ GeV and $g_X = 2.5 \times 10^{-4}$. 
	From the left to right and top to bottom panels, the mass splitting between the heavy hidden lepton generations 
	is set to be $\Delta m_{l^H} = 1$ GeV, $50$ GeV and $500$ GeV, respectively. 
	The dashed red, black and green lines represent the contributions from $\mathcal D$ boson, $\mathcal W'$ boson multiplied by $10^6$ 
	and $\mathcal H^{\pm}$ boson multiplied by $10^{66}$ respectively.
	The orange region is the excluded region at $90\%$ C.L. from MEG collaboration \cite{MEG:2016leq} and the dotted blue line indicates the future sensitivity from MEG II \cite{Meucci:2022qbh}. 
	}
\end{figure}

Fig.~\ref{fig:brmu2e-mlH} shows the branching ratio of $\mu \to e \gamma$ as a function of the heavy hidden lepton mass $m_{l^H}$. Here we fixed other parameters in the model to be 
$m_{h_2} = 292.50$ GeV, $m_{\mathcal D} = 766.07$ GeV, 
$m_{\mathcal H^{\pm}} = 848.13$ GeV, 
$m_{\mathcal W'} = 1.0$ GeV, $\theta_{1} = 0.030$ rad, 
$\theta_{2} = 0.056$ rad, 
$M_{X} = 1.96$ GeV and $g_X = 2.5 \times 10^{-4}$. We note that this benchmark point satisfies all current constraints mentioned above. 
The mass splitting $\Delta m_{l^H}$ is fixed to be $1$ GeV, $50$ GeV and $500$ GeV as respectively shown from the left to right and top to bottom panels in Fig.~\ref{fig:brmu2e-mlH}. 
From Fig.~\ref{fig:brmu2e-mlH} one can see that the contribution from the $\mathcal D$ boson diagram to the branching ratio of $\mu \to e \gamma$ is dominant. 
The SM $W$ boson contribution is suppressed by the sums over of $(\Delta m_{i1}/m_W)^4$ with $i = 2, 3$ and $\Delta m_{i1}^2$ is the mass difference between the neutrino generations. 
Using global fit values for $\Delta m_{i1}^2$ from \cite{Esteban:2020cvm}, one can obtain $\mathcal B (\mu \to e \gamma)_W \simeq 4.4 \times 10^{-55}$. 
The contribution from $\mathcal H^\pm$  is similarly suppressed by the mass of neutrinos,
whereas the contribution from $\mathcal W'$ is negligible due to 
the smallness of the gauge coupling $g_H$ that is $g_H \simeq 4.58 \times 10^{-4}$ for this benchmark point. 
For a fixed value of the mass splitting $\Delta m_{l^H}$, 
the total branching ratio of $\mu \to e \gamma$ decreases when $m_{l^H}$ increases. 
When $\Delta m_{l^H}$ increases, the branching ratio of $\mu \to e \gamma$ from $\mathcal D$ and $\mathcal W'$ bosons increase, while the contribution from charged Higgs is almost unchanged. 
For large values of $\Delta m_{l^H}$, the current limit from the MEG experiment can put a lower bound on $m_{l^H}$. A larger mass splitting $\Delta m_{l^H}$ requires a larger $m_{l^H}$. 
In particular, as shown on the top-right and bottom panel in Fig.~\ref{fig:brmu2e-mlH}, the heavy hidden lepton mass $m_{l^H}$ is required $\gtrsim$ 1 and 3 TeV for fixing $\Delta m_{l^H}$ at 50 and 500 GeV, respectively.
For small values of $\Delta m_{l^H}$, the branching ratio of $\mu \to e \gamma$ is suppressed and thus escaping the MEG constraint (see the top-left panel in Fig.~\ref{fig:brmu2e-mlH}). 
We note that for the degenerate mass case, {\it i.e.} $\Delta m_{l^H} = 0$, 
the contributions from new particles to the branching ratio of $\mu \to e \gamma$ vanishes. 
This is because, in this case, the form factors from $\mathcal D, \mathcal W'$ and $\mathcal H^{\pm}$ are proportional to the following factors~\footnote{
In general, we have $A_{ji}^{M/E} ( \mathcal{D \, {\rm or} \, W' \, {\rm or} \, H^\pm} ) \vert_{{\rm degenerate} \, f^H} = 0$ for $i \neq j$,
which is just the manifestation of the well-known GIM-mechanism in SM.}
\bea
A_{e\mu}^{M/E} (\mathcal D \, {\rm or} \,  \mathcal W' )\bigg\vert_{{\rm degenerate} \, l^H} &\sim& \sum_{k = 1}^{3} \left( {V_{l^H}} \right)_{2k}^{*} \left({V_{l^H} }\right)_{1k}   = 0\,, \\
A _{e\mu}^{M/E} (\mathcal H^\pm ) \bigg\vert_{{\rm degenerate} \, \nu^H} &\sim& \sum_{k = 1}^{3} \left( {V_{\nu^H}} \right)_{2k}^{*} \left({V_{\nu^H} }\right)_{1k}  = 0\,.
\eea

 \begin{figure}[tb]
         \centering
	\includegraphics[width=0.47\textwidth]{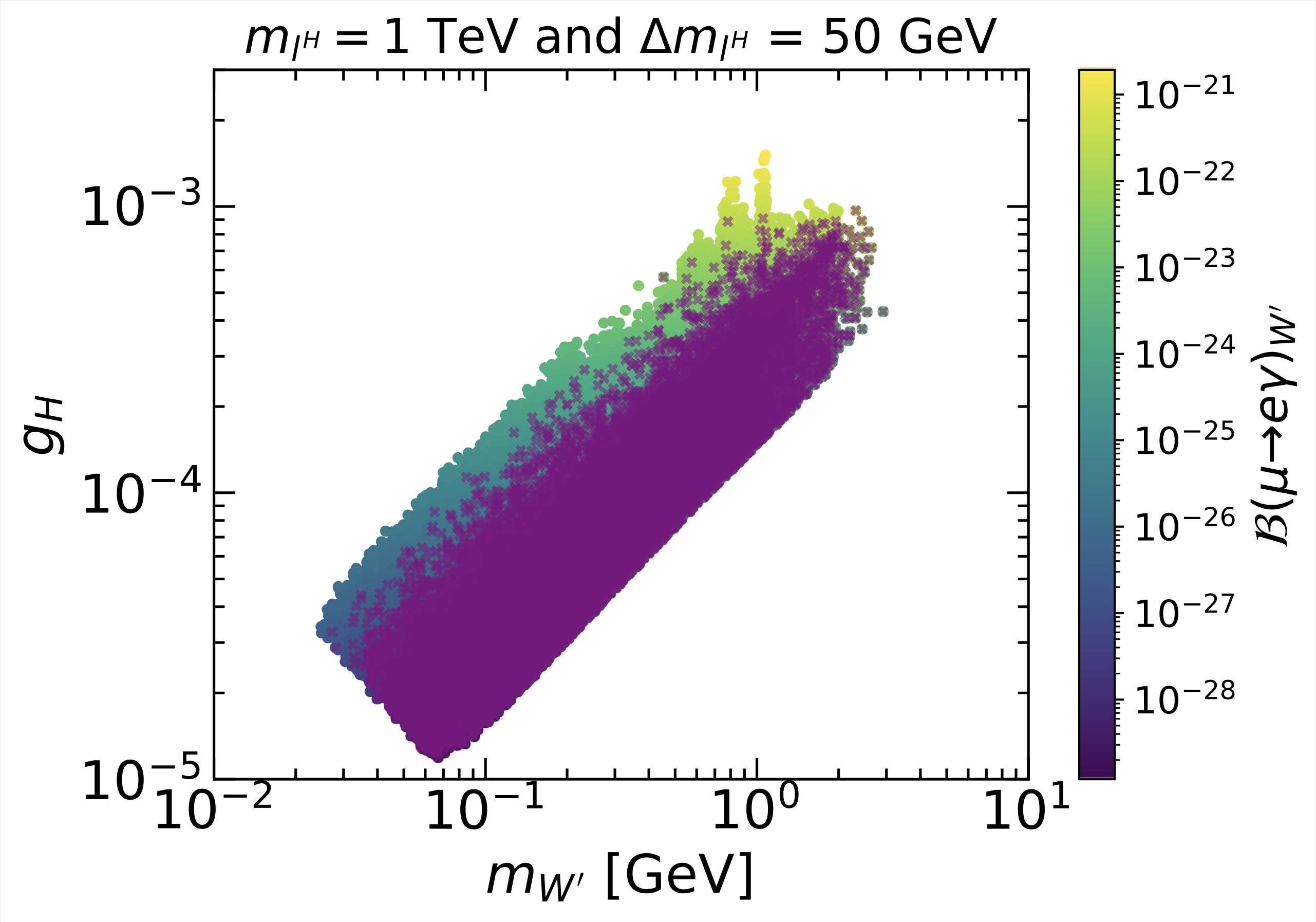}
	\includegraphics[width=0.47\textwidth]{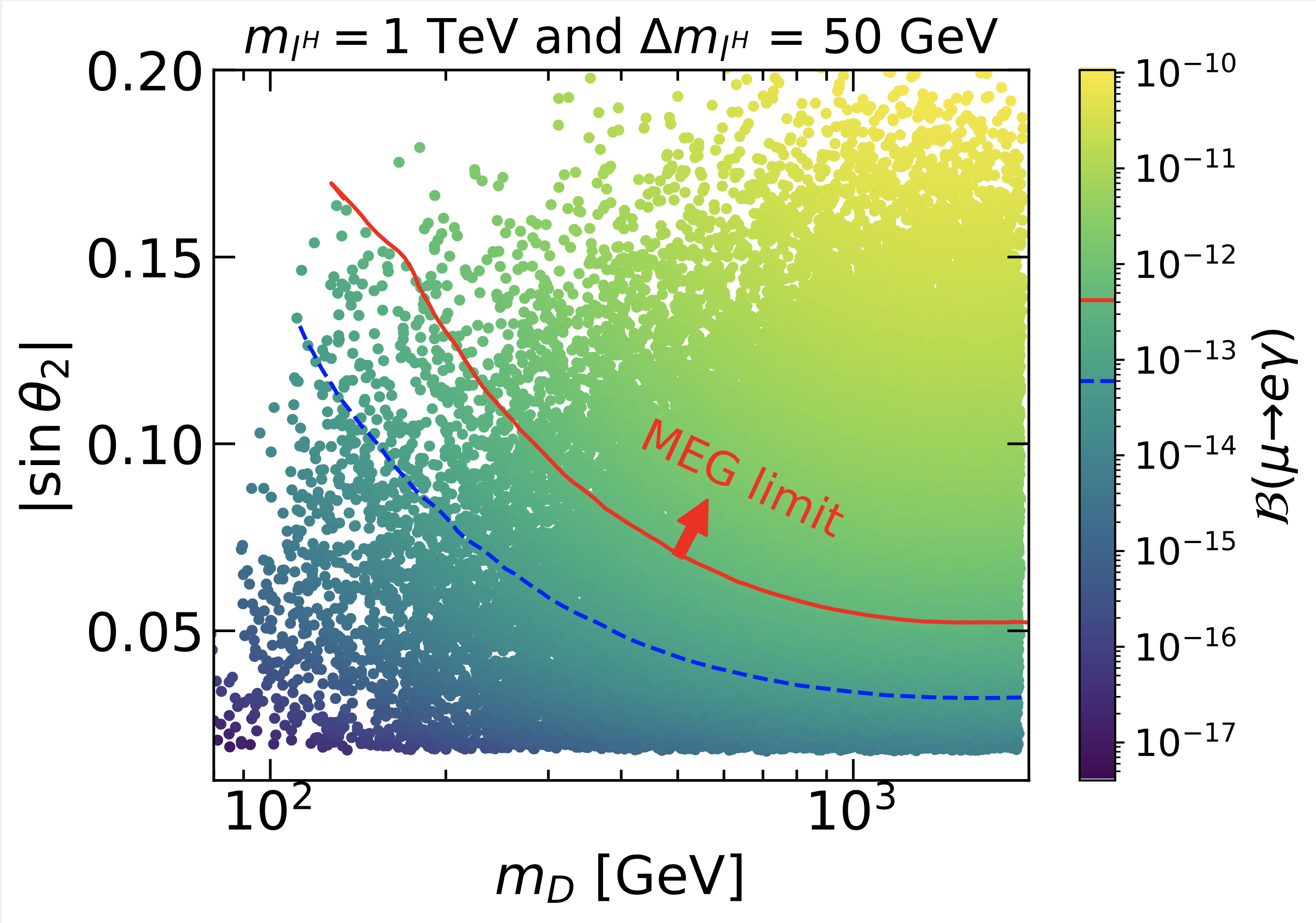}
	\caption{\label{fig:para-brm2e} Viable DM parameter points spanned in (${m_{\mathcal W'}, g_H}$) plane (left) and (${m_{\mathcal D}, |\sin \theta_2|}$) plane (right). 
	The color of circle points indicates the contribution to ${\cal B}(\mu \to e \gamma)$ from the $\mathcal W'$ diagram (left panel) and $\mathcal D$ diagram (right panel). 
	Here we fixed $m_{l^H} = 1$ TeV and $\Delta m_{l^H} = 50$ GeV. 
	The solid red and dashed blue lines on the right panel are the MEG limit and future sensitivity from MEG II \cite{Meucci:2022qbh}, respectively. 
	The crossed purple points on the left panel are allowed by the current limit from MEG \cite{MEG:2016leq} which are obtained from the bound on $|\sin \theta_2|$ (solid red on the right panel) due to the relation given in Eq.~(\ref{eq:gHtheta2relation}). 
	}
\end{figure}

Fig.~\ref{fig:para-brm2e} shows $2 \sigma$ favored parameter space for $\mathcal W'$ 
as a sub-GeV M candidate in the model. 
The data points are projected on (${m_{\mathcal W'}, g_H}$) plane (left panel) 
and (${m_{\cal D}, |\sin \theta_2|}$) plane (right panel). 
The colors of circle points in the left and right panels of Fig.~\ref{fig:para-brm2e} 
indicate the values of $\mathcal B (\mu \to e \gamma)$ 
calculated from $\mathcal W'$ diagram and $\mathcal D$ diagram, respectively. 
Here we fixed $m_{l^H} = 1$ TeV and $\Delta m_{l^H} = 50$ GeV. 
The contribution from the $\mathcal W'$ diagram to $\mathcal B (\mu \to e \gamma)$ is linearly proportional to the gauge coupling $g_H^2$.  
However due to the constraints from the dark matter direct detection and dark photon physics that required $g_H \lesssim 10^{-3}$ \cite{Ramos:2021omo, Ramos:2021txu, Tran:2022yrh}, 
the $\mathcal B (\mu \to e \gamma)$ from the $\mathcal W'$ diagram is suppressed. In particular, $\mathcal B (\mu \to e \gamma)_{\mathcal W'} \lesssim 10^{-21}$ as shown in the left panel of Fig.~\ref{fig:para-brm2e}.
On the other hand, the contribution to $\mathcal B (\mu \to e \gamma)$ from the $\mathcal D$ diagram is significant. 
The branching ratio is enhanced in the region of large mixing angle $\theta_2$ and heavy mass region of $\mathcal D$ boson. 
The current experimental data requires $0.018 \lesssim |\sin \theta_2| \lesssim 0.21$ for the $2\sigma$ favored region \cite{Tran:2022yrh}, which results $10^{-17} \lesssim \mathcal B (\mu \to e \gamma) \lesssim 10^{-10}$. 
We note that the ${\cal B} (\mu \to e \gamma)$ from the $\cal D$ boson diagram peaks at a certain value of $m_{\cal D}$ 
depending on the mass of heavy hidden leptons. 
For $m_{l^H} = 1$ TeV and $\Delta m_{l^H} = 50$ GeV, the peak is at $m_{\cal D} \sim 1.8$ TeV.
The current limit from the MEG experiment \cite{MEG:2016leq} can exclude a large portion of parameter space ($\sim 50\%$ of data points) in this enhanced region as shown by the red line on the right panel of Fig.~\ref{fig:para-brm2e}. 
The future sensitivity from MEG II \cite{Meucci:2022qbh}, as shown by the dashed blue line on the right panel of Fig.~\ref{fig:para-brm2e},
can probe lower values of the mixing angle $\theta_2$  and a smaller region of $\mathcal D$ boson mass in the model. 
We note that the upper bound on $|\sin \theta_2|$ from the MEG experiment can be translated into a bound on the DM mass $m_{\mathcal W'}$ 
and the gauge coupling $g_H$ 
due to the following relation 
\be
\label{eq:gHtheta2relation}
g_H  = \frac{2 \,m_{W'}}{v} \times
\begin{cases}
\begin{matrix}
 |\sin \theta_2|  \, , &&  {\rm for} \; \theta_2 > 0 \; , \\
 |\cos \theta_2| \, , &&  {\rm for} \; \theta_2 \leq 0 \; ,
\end{matrix}
\end{cases} 
\ee
which can be derived from Eqs.~(\ref{mwprime}) and (\ref{theta2def}).
The allowed points after taking into account the MEG constraint projected on ($m_{\mathcal W'}, g_H$) plane is shown as the crossed purple points in the left panel of Fig.~\ref{fig:para-brm2e}.

\begin{figure}[tb]
         \centering
	\includegraphics[width=0.55\textwidth]{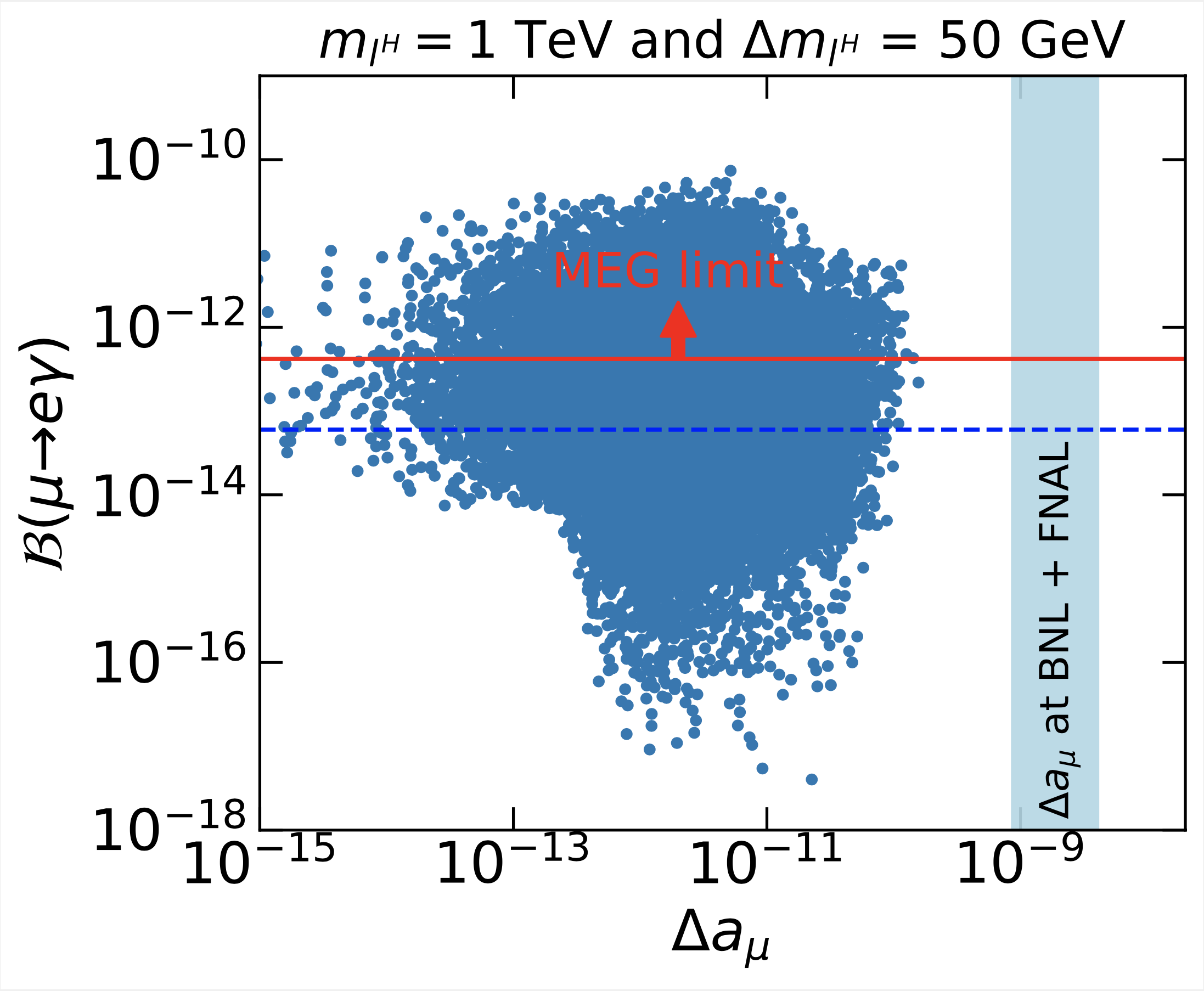}
	\caption{\label{fig:deltamu-brm2e} Viable DM parameter points spanned in the plane of the total branching ratio of $\mu \to e \gamma$ and muon anomalous magnetic dipole moment $\Delta a_\mu$ . 
	Here we fixed $m_{l^H} = 1$ TeV and $\Delta m_{l^H} = 50$ GeV. The solid red and dashed blue lines are the current limit from MEG \cite{MEG:2016leq} and future sensitivity from MEG II \cite{Meucci:2022qbh}, respectively. The shaded light blue band represents 
	the $2\sigma$ region of $\Delta a_\mu$ measured at BNL \cite{Muong-2:2002wip,Muong-2:2004fok,Muong-2:2006rrc} and FNAL \cite{Muong-2:2021ojo}. 
	}
\end{figure}

In Fig.~\ref{fig:deltamu-brm2e}, we show the $2 \sigma$ favored parameter space on the plane of 
the total branching ratio of $\mu \to e \gamma$ 
and muon anomalous magnetic dipole moment $\Delta a_\mu$. 
As mentioned above, the main contribution to the $\mathcal B (\mu \to e \gamma)$ is from the $\mathcal D$ boson diagram 
and a large portion of parameter space can be excluded by the current MEG experiment. 
On the other hand, the main contributions to $\Delta a_\mu$ in the model are from $\mathcal D$ and $Z_{2,3}$ diagrams. 
The contribution from the $\mathcal W^\prime$ diagram gives a negative value for $\Delta a_\mu$, 
whereas the neutral Higgs and charged Higgs contributions are both suppressed 
for the current viable parameter space in the model. 
One can see in Fig.~\ref{fig:deltamu-brm2e}, the total contribution to $\Delta a_\mu$ 
is not reaching the $2\sigma$ region (shaded light blue) for the muon anomalous magnetic dipole moment measured at BNL \cite{Muong-2:2002wip,Muong-2:2004fok,Muong-2:2006rrc} and FNAL \cite{Muong-2:2021ojo}. 
We expect the $\Delta a_\mu$ can be enhanced in higher loop diagrams such as the two-loop Barr-Zee mechanism~\cite{Barr:1990um}. 
Calculation of these two-loop Barr-Zee contributions is thus highly desirable but nevertheless beyond the scope of this study. We hope to return to this issue in the future. 

\begin{figure}[tb]
         \centering
	\includegraphics[width=0.55\textwidth]{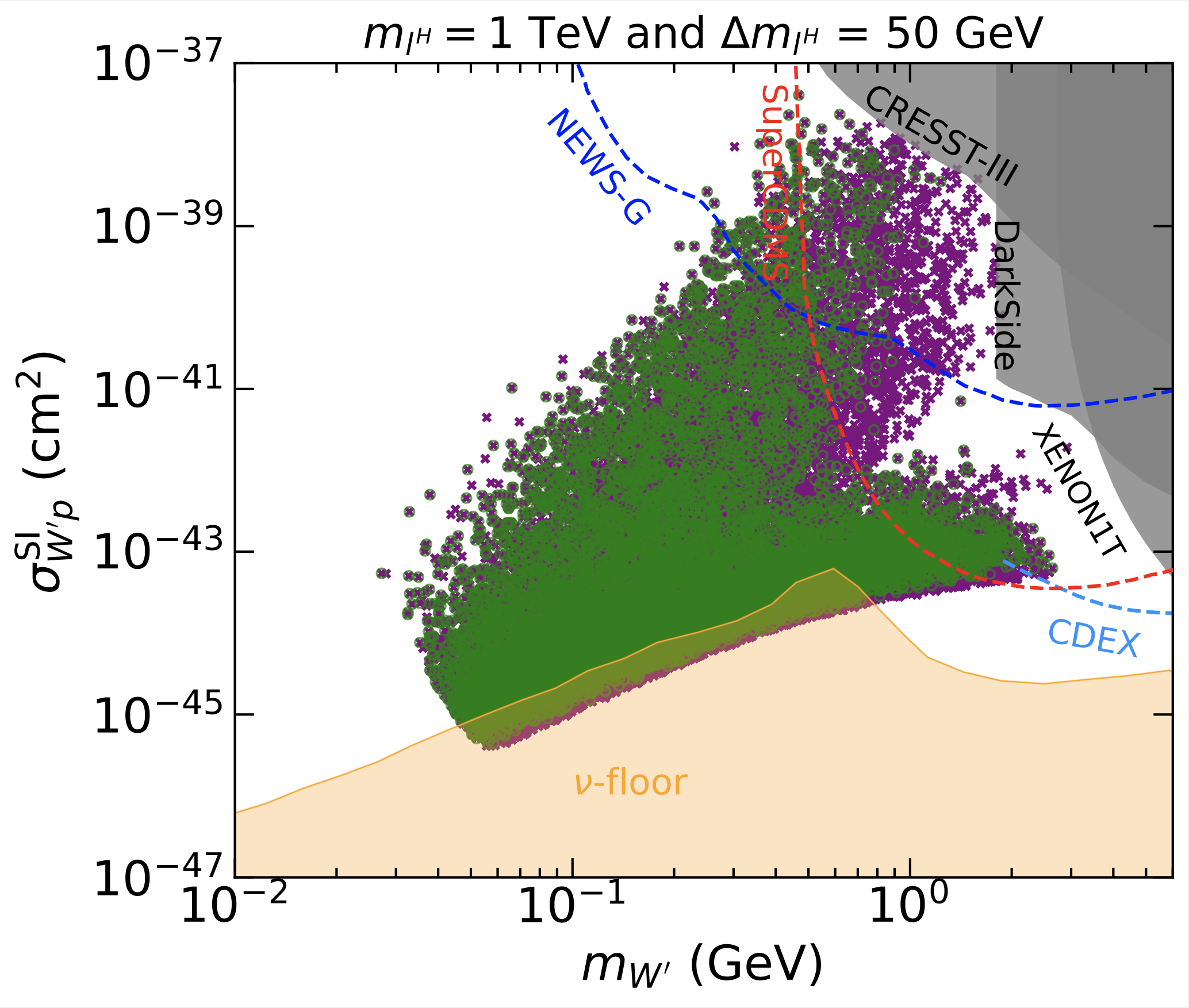}
	\caption{\label{fig:mWp-DMxsec} Favored data projected on the plane of the DM mass and spin independent DM-proton scattering cross section. 
	Here we fixed $m_{l^H} = 1$ TeV and $\Delta m_{l^H} = 50$ GeV. The crossed purple points indicate the data satisfied the MEG constraint \cite{MEG:2016leq}, while the circle green points indicate the data that can be probed by future experiment from MEG II \cite{Meucci:2022qbh}. 
	The gray regions are the exclusion from CRESST-III~\cite{Angloher:2017sxg}, DarkSide-50 \cite{Agnes:2018ves} and XENON1T \cite{Aprile:2019xxb} experiments. The dashed blue, red and light blue lines represent the future sensitivities from DM direct detection experiments 
	at NEWS-G \cite{Battaglieri:2017aum}, SuperCDMS \cite{Agnese:2016cpb} and CDEX \cite{Ma:2017nhc}, respectively. Orange region is the neutrino floor background. 
	}
\end{figure}

Fig.~\ref{fig:mWp-DMxsec} shows the allowed data points by the current MEG constraints (crossed purple) 
and the points that can be probed by future experiment from MEG II (circle green) on the DM direct detection plane. 
The predicted DM mass range in $2\sigma$ favored region is $\sim (0.02 - 3)$ GeV. The future sensitivity from MEG II can probe almost the entire viable range of DM mass. 
Interestingly, some data points with the DM mass at around $1$ GeV can be probed complementarily by various future DM direct detection experiments including NEWS-G \cite{Battaglieri:2017aum}, SuperCDMS \cite{Agnese:2016cpb} and CDEX \cite{Ma:2017nhc}.

\section{Conclusion} 
We computed the one-loop radiative decay rates for 
the charged lepton flavor violation processes $l_{i} \to l_{j} \gamma$, the anomalous magnetic dipole moment $\Delta a_{l_i}$ 
and the electric dipole moment $d_{l_i}$ of the charged lepton $l_i$ at one-loop level in a minimal G2HDM. 
Besides the contribution from the SM $W$ boson, the charged lepton flavor violation processes occurred at one-loop in G2HDM 
due to the new interactions of SM charged leptons with two $h$-parity odd particles -- heavy hidden leptons ($l^H$ or $\nu^H$) 
and  hidden dark scalars or gauge bosons ($\mathcal D$, $\mathcal H^{\pm}$,$\mathcal W'$). 
The contributions from these new interactions vanish when the heavy hidden lepton masses among generations degenerate. 

We analyzed the $\mu \to e \gamma$ process and $\Delta a_{\mu}$ 
using a parameter space that favors a sub-GeV non-abelian vector dark matter $\mathcal W^{\prime (p,m)}$ in the model. 
The scan data is adapted from Ref.~\cite{Tran:2022yrh} and they satisfy various constraints
including the theoretical constraints on the scalar potential, 
the Higgs signal strength measurements from the LHC, 
the dark photon physics, 
the electroweak precision measurements including the recent $W$ boson mass measurement at the CDF II,
DM relic density measured from Planck collaboration,
and from DM searches including the DM direct detections and the Higgs invisible width from the LHC.

We found that the branching ratio of $\mu \to e \gamma$ 
is significantly dependent on the heavy hidden lepton mass and the mass splitting between its generations.
In particular, a heavier hidden lepton mass results in a smaller branching ratio of $\mu \to e \gamma$ 
while a larger mass splitting gives a larger branching ratio as shown in Fig.~\ref{fig:brmu2e-mlH}. 

Among the new contributions to $\mu \to e \gamma$ in the model, the contribution 
from $\mathcal D$ boson diagram is dominant. 
The branching ratio can be enhanced in the heavy $\mathcal D$ boson mass region and the region of large mixing angle $\theta_2$ between 
two odd $h$-parity bosons, $\mathcal H^0_2$ and $G_H^m$, which compose $\mathcal D$.
The current constraint on the cLFV process from MEG can exclude a significant portion of the parameter space in the favored region obtained in previous studies. 
Although the contribution from the DM candidate $\mathcal W'$ to the branching ratio of $\mu \to e \gamma$ is suppressed 
due to the smallness of new gauge coupling $g_H$, 
the DM parameters can be affected indirectly by the cLFV processes
due to the relation between parameters in the model, especially the relation (\ref{eq:gHtheta2relation}) between the mixing angle
$\theta_2$, new gauge coupling $g_H$ and DM mass $m_{\mathcal W'}$. 
We found that the future measurement at MEG II can probe almost the entire viable range of the DM mass
which is $\sim (0.02 - 3)$ GeV and interestingly, the region at the DM mass around $1$ GeV can be also probed by future DM direct detection experiments such as NEWS-G, SuperCDMS and CDEX.

In the viable DM parameter space, the total one-loop level contribution to $\Delta a_{\mu}$ in the model is not big enough to explain the $4.2 \sigma$ level discrepancy between 
the theoretical value and the experimental results measured at the BNL and FNAL.
We expect an enhanced contribution to $\Delta a_{\mu}$ at higher loop corrections, such as the Barr-Zee two-loop mechanism~\cite{Barr:1990um} for the neutrino magnetic dipole moments,
can be anticipated to address the muon anomaly in the model. 

In Appendix A, we also showed that the electric dipole moment of charged lepton vanishes at one-loop in G2HDM. 
This is due to (1) the lack of CP violating phases (or in general imaginary parts) in products of generic but related complex vector and axial vector gauge couplings or 
scalar and pseudoscalar Yukawa couplings, and (2) vanishing combinations of Feynman loop integrals.
Same conclusion can be drawn for the SM quarks in the model.
Thus it is also interesting to investigate if the two-loop mechanisms like the Weinberg three-gluon operator~\cite{Weinberg:1989dx,Braaten:1990gq,Weinberg:1990me} 
 for the gluon chromo-electric dipole moment~\cite{Braaten:1990zt} and the Barr-Zee diagrams for the charged leptons~\cite{Barr:1990vd} can generate a non-vanishing result for the electric dipole moments for the
neutron and SM leptons respectively in G2HDM. 
For these two-loop calculations, we will reserve them for future tasks.
 
 \newpage 
 \section*{Acknowledgments}

We would like to thank Michael J. Ramsey-Musolf for encouragements and useful discussions .
This work is supported in part by the Ministry of Science and Technology (MOST) of Taiwan under Grant No 111-2112-M-001-035  
(TCY) and by National Natural Science Foundation of China under Grant No. 19Z103010239 (VQT).
VQT would like to thank the High Energy Theory Group at the Institute of Physics, Academia Sinica, Taiwan for its hospitality.


\renewcommand{\theequation}{A.\arabic{equation}}
\setcounter{equation}{0}
\section*{Appendix A. One-Loop Analytical Formulae of $A^M_{ji}$ and $A^E_{ji}$}
\label{FormFactors}

In this Appendix, we present the one-loop expressions for the transition magnetic and electric dipole form factors 
$A^M_{ji}$ and $A^E_{ji}$
from the six different contributions in G2HDM. For the gauge particle loops of $W^\pm$, $\{Z_n\}$ and $\mathcal W^{\prime (p,m)}$, we use unitary gauge in the computation.
For the computation of $W^\pm$ loop in the 't Hooft-Feynman gauge, see Appendix B.
\footnote{The issue of gauge fixings in the model has been studied as given in an Appendix in Ref. \cite{Ramos:2021txu}. }
For convenience, we define $z=1-x-y$ in what follows.

\subsection{$W$ contribution --  Left diagram in Fig.~(\ref{fig:li-lj-gamma-SM-like})}

The transition magnetic and electric dipole form factors are given by
\begin{align}
A^M_{ji} \left( W \right) & =   + \frac{1}{ 8 \pi^2 } \left( \frac{g}{2 \sqrt 2} \right)^2  
\sum_k \left( V_{\rm PMNS} \right)^*_{kj} \left( V_{\rm PMNS}\right)_{ki} \nonumber \\ 
& \;\;\;\;\;\;\;\;\;\;\;\; \times \left[ \mathcal I  \left( m_i , m_j, m_{\nu_k}, m_W \right)  +
\mathcal I  \left( m_i , m_j, - m_{\nu_k}, m_W \right)  \right] \; , \\
\label{AEW}
A^E_{ji} \left( W \right) & =   -  \frac{i}{ 8 \pi^2 }  \left( \frac{g}{2 \sqrt 2} \right)^2  
\sum_k \left( V_{\rm PMNS} \right)^*_{kj} \left( V_{\rm PMNS}\right)_{ki} \nonumber \\ 
& \;\;\;\;\;\;\;\;\;\;\;\; \times \left[ \mathcal I  \left( m_i , -m_j, m_{\nu_k}, m_W \right)  +
\mathcal I  \left( m_i , -m_j, - m_{\nu_k}, m_W \right)  \right] 
\; ,
\end{align}
respectively where the loop integral $\mathcal I  \left( m_i , m_j , m_k , m_X \right) $ is 
\begin{align}
\label{Integral-I}
\mathcal I & \left( m_i , m_j , m_k , m_X \right) \nonumber \\ 
& = \int_0^1 {\rm d} x \int_0^{1-x} {\rm d} y \Biggl\{ \frac{1} { -x z m_i^2 - x y m_j^2 + x m_k^2 + (1 - x ) m_X^2 } \Biggr. \nonumber \\
& \;\;\;\; \times \Biggl[ \left( \bigl( y + 2 z \left( 1 - x \right) \bigr) + \bigl( z + 2 y \left( 1 - x \right) \bigr) \frac{ m_j }{ m_i } - \, 3  \left( 1 - x \right)  \frac{ m_k }{ m_i }  \right)
\Biggr. \nonumber \\
& \;\;\;\;\;\;\;\;\; + \Biggl. \frac{m_i^2}{m^2_{X}}  x^2 \left(   z^2  + y^2 \frac{m^3_j }{m_i^3} 
+ y z \frac{m_j}{m_i} \left( 1 + \frac{m_j}{m_i} \right)  - \frac{m_j m_k }{m_i^2} \right)  \Biggr]  \nonumber \\
&  \;\;\;\; + \frac{1}{m^2_X} \left( x (1-z)  + y + \bigl( x \left(1 - y \right) + z \bigr) \frac{m_j}{m_i}  -  \frac{m_k}{m_i} \right) \nonumber \\
&  \;\;\;\; + \frac{1}{m^2_X} \left(  2 - x \left( 3 - 4 z \right) - 3 y - z + \bigl(  2 - x \left( 3 - 4 y \right) - y - 3 z \bigr) \frac{ m_j }{ m_i }  \right) \nonumber \\
& \;\;\;\;\;\;\;\;\; \Biggl.  \times  \log \left( \frac{m^2_X} {-x z m_i^2 - x y m_j^2 + x m_k^2 + (1 - x ) m_X^2} \right)  \Biggr\} \; .
\end{align} 
We note that this integral $\mathcal I$ is for the diagram with two internal charged vector bosons $X$ coupled to the external photon 
computed using the unitary gauge. The third line of Eq.~(\ref{Integral-I}) comes from the product of the transverse pieces of the two vector boson propagators, while all the remaining terms are 
due to the product of the transverse and longitudinal pieces of these two propagators. The product of longitudinal pieces do not give rise to the contributions for the transition magnetic and 
electric dipole form factors. 
Our integral $\mathcal I$ is denoted as $I^{++}_{k,3}$ in Eq.~(A.5) of~\cite{Lindner:2016bgg}.
Except for the fourth line of Eq.~(\ref{Integral-I}), our formula agrees~\footnote{Note that there are also a couple of trivial typos 
in the coefficients of the log term of $I^{++}_{k,3}$ in Eq.~(A.5) of~\cite{Lindner:2016bgg}.}. 
The difference between our result of Eq.~(\ref{Integral-I}) and Eq.~(A.5) of~\cite{Lindner:2016bgg} is
\begin{align}
{\rm Diff} & = \frac{ \left( m_i - m_j \right)^2 }{m_X^2}  \int_0^1 {\rm d} x \int_0^{1-x} {\rm d} y 
\frac{ x y z \left( x + y + \left( 1 - y \right) \frac{m_j}{m_i} - \frac{ m_k }{ m_i } \right) } { -x z m_i^2 - x y m_j^2 + x m_k^2 + (1 - x ) m_X^2 }  \; .
\end{align}
Since the difference disappears in the case of $m_i = m_j$, one can't use the known result of charged lepton anomaly~\cite{Leveille:1977rc} to discriminate the two answers.
However, see Appendix B.

For $i=j$, there is no CP violating phase arise from 
the product $ \vert \left( V_{\rm PMNS}\right)_{ki} \vert^2$ in (\ref{AEW}), which implies the electric dipole moment $d_{l_i}$ should be vanishing from the 
one-loop $W$ diagram in SM. Indeed the sum of the two integrals in (\ref{AEW}) vanishes when $m_i=m_j$!

\subsection{$\left\{ Z_n \right\}$ contribution -- Middle diagram in Fig.~(\ref{fig:li-lj-gamma-SM-like})}

The transition magnetic dipole form factor is given by 
\begin{align}
\label{AMZ}
A^M_{ji} \left( \left\{ Z_n \right\} \right) & =
+\frac{\delta_{ij}}{ 8 \pi^2 }  \sum_n  \left[  \left( C_{V \, n} \right)^2 \mathcal J  \left( m_i , m_i,  m_i, m_{Z_n} \right)  \right. \nonumber \\
& \quad\quad\quad\quad\quad  \left. + \left( C_{A \, n} \right)^2  \mathcal J  \left( m_i , m_i, - m_i, m_{Z_n} \right)  \right] \; , 
\end{align}
where
\begin{align}
\label{Integral-J}
\mathcal J & \left( m_i , m_j , m_k , m_X \right) \nonumber \\ 
& = - \int_0^1 {\rm d} x \int_0^{1-x} {\rm d} y \Biggl\{ \frac{1} { -x z m_i^2 - x y m_j^2 + (1-x) m_k^2 + x m_X^2 } \Biggr. \nonumber \\
& \;\;\;\; \times \Biggl\lceil  2 x \left( (1-z) + (1-y) \frac{m_j}{m_i} - 2  \frac{m_k}{m_i} \right) \Biggr. \nonumber \\
& \;\;\;\;\;\;\;\;\; + \frac{m_i^2}{m^2_X} \Biggl( (1-x) \left( \frac{m_j}{m_i} - \frac{m_k}{m_i} \right) 
\left( z + y \frac{m_j}{m_i} \right) \left( 1 - \frac{m_k}{m_i} \right) \Biggr.  \nonumber \\
& \;\;\;\;\;\;\;\;\;\;\;\;\;\;\;\;\;\;\;\;\;\; - z \left( \frac{m_j}{m_i} - \frac{m_k}{m_i} \right) \left(  (1-x (1 - z) ) + x y \frac{m_j^2}{m_i^2} \right) \nonumber \\
& \Biggl. \Biggl. \;\;\;\;\;\;\;\;\;\;\;\;\;\;\;\;\;\;\;\;\;\;  - \, y \left( 1 - \frac{m_k}{m_i} \right) \left( xz + (1-x (1 - y) ) \frac{m_j^2}{m_i^2} \right) \Biggr) \Biggr\rfloor \nonumber \\
&  \;\;\;\;  +  \frac{1}{m^2_X} \left( y  + z \frac{m_j}{m_i } -  \left( 1 - x  \right) \frac{m_k}{m_i} \right) \nonumber \\
&  \;\;\;\; + \frac{1}{m^2_X} \left( (1 - 3y )  + (1 - 3 z ) \frac{m_j}{m_i} +  \left( 1 - 3 \, x \right) \frac{m_k}{m_i} \right) \nonumber \\
& \;\;\;\;\;\;\;\;\; \Biggl. \times \log \left( \frac{m^2_X} {-x z m_i^2 - x y m_j^2 + (1-x) m_k^2 + x m_X^2} \right)  \Biggr\} \; .
\end{align} 
We note that this integral of $\mathcal J$ is for the diagram with one internal neutral gauge boson $X$ exchange computed using the unitary gauge. 
The third line of Eq.~(\ref{Integral-J}) comes from the transverse piece of the vector boson propagator, while the remaining terms come entirely from the 
longitudinal piece of the propagator.  Our loop integral $(- \mathcal J)$ corresponds to $I^{++}_{k,4}$ in Eq.~(A.6) of~\cite{Lindner:2016bgg}.~\footnote{
We note that in the fourth line of our Eq.~(\ref{Integral-J}), instead of the factor of $(1-x)$, Eq.~(A.6) of~\cite{Lindner:2016bgg} got
$(x-1)$.}
Using our expression of $\mathcal J$ in Eq.~(\ref{Integral-J}) for the equal mass case of $m_i = m_j = m_\mu$ and setting $m_k = m_F$, 
one can easily reproduce the well-known expression of muon anomaly for a neutral gauge boson $X$ with a general gauge coupling of a muon and another fermion $F$ first obtained in~\cite{Leveille:1977rc}.

For the transition electric dipole from factor, one finds
\begin{align}
\label{AEZ}
A^E_{ji} \left( \left\{ Z_n \right\} \right) & =   0  \;,
\end{align}
which implies $d_{l_i} (\{ Z_n \} )=0$.

\subsection{$\left\{ h_n \right\}$ contribution -- Right diagram in Fig.~(\ref{fig:li-lj-gamma-SM-like})}
The transition magnetic dipole form factor is
\begin{align}
A^M_{ji} \left( \left\{ h_n \right\} \right) & = 
 \frac{\delta_{ij}}{ \pi^2 } \frac{m_i^2}{v^2}  \sum_n  \left(  O^H \right)_{1n}^2 \mathcal K  \left( m_i , m_i,  m_i, m_{h_n} \right)  \; , 
\end{align}
with
\begin{align}
\label{Integral-K}
\mathcal K \left( m_i , m_j , m_k , m_X \right) & = 
\int_0^1 {\rm d} x \int_0^{1-x} {\rm d} y  \nonumber \\
& \;\;\;\; \times \left[ \frac{ x \left( y +  z \frac{m_j}{m_i} \right) +  (1-x) \frac{m_k}{m_i} } { -x y m_i^2 - xz m_j^2  + (1-x) m_k^2 + x m_X^2 } \right] \; .
\end{align} 
This loop integral $\mathcal K$ is the same as $I^{++}_{k,1}$ in Eq.~(A.1) of~\cite{Lindner:2016bgg}. 

As in the $\{ Z_n \}$ case, one finds that the transition electric dipole form factor vanishes
\beq
\label{AEh}
A^E_{ji} \left( \left\{ h_n \right\} \right)  =   0 \;, 
\eeq
which implies  $d_{l_i}(\{ h_n \})=0$ as well.

\subsection{$\mathcal D$ contribution -- Left diagram in Fig.~(\ref{fig:li-lj-gamma-G2HDM})}
The transition magnetic and electric dipole form factors are given by
\begin{align}
\label{AMD}
A^M_{ji} \left( \mathcal D \right) & =  
\frac{1}{ 8 \pi^2  }  \left[ \sum_k y^{\mathcal D \, *}_{S \, kj} y^{\mathcal D}_{S \, ki} \mathcal K  \left( m_i , m_j, m_{l^H_k}, m_{\mathcal D} \right)  
\right. \nonumber \\
& \;\;\;\;\;\;\;\;\;\;\;\;\;\;  +  \left. \sum_k y^{\mathcal D \, *}_{P \, kj} y^{\mathcal D}_{P \, ki} \mathcal K  \left( m_i , m_j, -m_{l^H_k}, m_{\mathcal D} \right) \right] \; , \\
\label{AED}
A^E_{ji} \left( \mathcal D \right) & =   
 \frac{i}{ 8 \pi^2 } \left[ \sum_k y^{\mathcal D \, *}_{P \, kj} y^{\mathcal D}_{S \, ki} \mathcal K  \left( m_i , -m_j, m_{l^H_k}, m_{\mathcal D} \right) 
\right. \nonumber \\
&  \;\;\;\;\;\;\;\;\;\;\;\;\;\; + \left. \sum_k y^{\mathcal D \, *}_{S \, kj} y^{\mathcal D}_{P \, ki} \mathcal K  \left( m_i , -m_j, -m_{l^H_k}, m_{\mathcal D} \right) \right] \; ,
\end{align}
where the summation is over all the heavy hidden charged leptons $l^H_k$ running inside the loop. $\mathcal K$ is defined already in (\ref{Integral-K}).

For $i=j$,   (\ref{AMD}) reduces to
\begin{align}
A^M_{ii} \left( \mathcal D \right) & =    \frac{1}{ 8 \pi^2 }  \left[
 \sum_k \vert  y^{\mathcal D}_{S \, ki} \vert^2 \mathcal K  \left( m_i , m_i, m_{l^H_k}, m_{\mathcal D} \right) \right. \nonumber \\
& \;\;\;\;\;\;\;\;\;\;\;\;\;\;\ + \left. \sum_k  \vert y^{\mathcal D}_{P \, ki}  \vert^2 \mathcal K  \left( m_i , m_i, -m_{l^H_k}, m_{\mathcal D} \right)  \right] \; ,
\end{align}
with
\bea
 \vert  y^{\mathcal D}_{(S,P) \, ki} \vert^2  & = \frac{1}{2} \vert \left( V^H_l \right)_{ik} \vert^2 \left( \frac{m_{l_i}^2}{v^2} \cos^2\theta_2  + \frac{m_{l^H_k}^2}{v_\Phi^2} \sin^2\theta_2 \pm
 \frac{m_{l_i} m_{l^H_k}} {v v_\Phi} \sin 2 \theta_2 \right)  \;.
\eea
We note that the possible new CP violating phase in $V^H_l$ is cancelled out in $\vert y^{\mathcal D}_{(S, P) \, ki} \vert^2$.

On the other hand, for $i=j$,  (\ref{AED}) reduces to
\begin{align}
A^E_{ii} \left( \mathcal D \right) & =    \frac{i}{ 8 \pi^2 }  
 \sum_k {\rm Re} \left( y^{\mathcal D \, *}_{P \, ki} y^{\mathcal D}_{S \, ki}  \right) 
\left[ \mathcal K  \left( m_i , -m_i, m_{l^H_k}, m_{\mathcal D} \right) 
 + \mathcal K  \left( m_i , -m_i, -m_{l^H_k}, m_{\mathcal D} \right) \right]  \nonumber \\
&   -  \frac{1}{8 \pi^2 }   \sum_k {\rm Im} \left( y^{\mathcal D \, *}_{P \, ki} y^{\mathcal D}_{S \, ki}  \right) 
\left[ \mathcal K  \left( m_i , -m_i, m_{l^H_k}, m_{\mathcal D} \right) 
 - \mathcal K  \left( m_i , -m_i, -m_{l^H_k}, m_{\mathcal D} \right) \right] \nonumber \\
& =  0 \; ,
\end{align}
due to the fact that $\mathcal K  \left( m_i , -m_i, m_{l^H_k}, m_{\mathcal D} \right) 
 + \mathcal K  \left( m_i , -m_i, -m_{l^H_k}, m_{\mathcal D} \right) = 0$ and from (\ref{yPsyS}) we have
${\rm Im} \left( y^{\mathcal D \, *}_{P \, ki} y^{\mathcal D}_{S \, ki}  \right) =0$. 

\subsection{$\mathcal H^\pm$ contribution -- Middle diagram in Fig.~(\ref{fig:li-lj-gamma-G2HDM})}
The transition magnetic and electric dipole form factors are 
\begin{align}
\label{AMH}
A^M_{ji} \left( \mathcal H \right) & =  + \frac{1}{8 \pi^2 } \sum_k y^{\mathcal H \, *}_{S \, kj} y^{\mathcal H}_{S \, ki} 
\left[ 
\mathcal L   \left( m_i , m_j, m_{\nu^H_k}, m_{\mathcal H} \right) \right.  \nonumber \\
& \quad \quad \quad \quad \quad \quad \quad \quad \quad \quad \left. +  \mathcal L  \left( m_i , m_j, -m_{\nu^H_k}, m_{\mathcal H} \right) 
\right]  \; , \\
\label{AEH}
A^E_{ji} \left( \mathcal H \right) & =  - \frac{i}{ 8 \pi^2 } \sum_k y^{\mathcal H \, *}_{S \, kj} y^{\mathcal H}_{S \, ki} 
\left[ 
\mathcal L   \left( m_i , -m_j, m_{\nu^H_k}, m_{\mathcal H} \right) \right. \nonumber \\
& \quad \quad \quad \quad \quad \quad \quad \quad \quad \quad \left. +  \mathcal L  \left( m_i , -m_j, -m_{\nu^H_k}, m_{\mathcal H} \right) 
\right] \; ,
\end{align}
where we have used $y^{\mathcal H}_{P \, kj}  = -y^{\mathcal H}_{S \, kj}$ from (\ref{YukawaH}) and the summation is over all the heavy hidden neutrinos $\nu^H_k$ running inside the loop. 
The loop integral $\mathcal L$ is given by 
\begin{align}
\label{Integral-L}
\mathcal L \left( m_i , m_j , m_k , m_X \right) & = -
\int_0^1 {\rm d} x \int_0^{1-x} {\rm d} y \nonumber \\
& \;\;\;\; \times \left[ \frac{ x \left( y +  z \frac{m_j}{m_i} +  \frac{m_k}{m_i} \right) } { -x y m_i^2 - xz m_j^2  + x m_k^2 + (1-x) m_X^2 } \right] \; .
\end{align} 
Our loop integral $(-\mathcal L)$ is the same as $I^{++}_{k,2}$ in Eq.~(A.2) of~\cite{Lindner:2016bgg}.

For $i=j$, each term in both $A^M_{ii}(\mathcal H)$ and $A^E_{ii}(\mathcal H)$ is proportional to 
\beq
\vert y^{\mathcal H }_{S \, ki} \vert^2 = \frac{ \vert \left( V^{H \, \dagger}_\nu M_\nu V_{\rm PMNS} \right)_{ki} \vert^2 } {2 v^2} \; ,
\eeq
which is real but may contain CP-violating phases from $V_{\rm PMNS}$ and $V^H_\nu$. 
The effects from these CP-violating phases in $\Delta a_{l_i}(\mathcal H^\pm)$ are small due to the suppression from the neutrino masses.
The important role of CP violating phases 
in the muon anomaly in MSSM coming from the charginos and neutrinos sectors has been emphasized previously in~\cite{Ibrahim:1999hh,Ibrahim:1999aj,Ibrahim:2001ym}. 
The electric dipole moment $d_{l_i}(\mathcal H^\pm)$ should be vanishing since the sum of the two integrals in (\ref{AEH}) vanishes when $m_i=m_j$!

\subsection{$\mathcal W^{\prime (p,m)}$ contribution -- Right diagram in Fig.~(\ref{fig:li-lj-gamma-G2HDM})}
The transition magnetic and electric dipole form factors are 
\begin{align}
A^M_{ji} \left( \mathcal W' \right) & =  + \frac{1}{ 8 \pi^2 }  \sum_k g^{\mathcal W' \, *}_{V \, kj} g^{\mathcal W'}_{V \, ki} 
\left[ 
\mathcal J  \left( m_i , m_j, m_{l^H_k}, m_{\mathcal W'} \right) \right. \nonumber \\
& \quad \quad \quad \quad \quad \quad \quad \quad \quad \quad \left. +  \mathcal J  \left( m_i , m_j, -m_{l^H_k}, m_{\mathcal W'} \right) 
\right]  \; , \\
\label{AEWp}
A^E_{ji} \left( \mathcal W' \right) & =   + \frac{i}{ 8 \pi^2 } \sum_k g^{\mathcal W' \, *}_{V \, kj} g^{\mathcal W'}_{V \, ki} 
\left[ 
\mathcal J   \left( m_i , -m_j, m_{l^H_k}, m_{\mathcal W'} \right) \right. \nonumber \\
& \quad \quad \quad \quad \quad \quad \quad \quad \quad \quad \left. +  \mathcal J  \left( m_i , -m_j, -m_{l^H_k}, m_{\mathcal W'} \right) 
\right] \; ,
\end{align}
where we have used $g^{\mathcal W'}_{A \, kj} = g^{\mathcal W'}_{V \, kj} $ from (\ref{gaugecouplingWp}) 
and  the summation is over all the heavy hidden charged leptons $l^H_k$ running inside the loop.
The loop integral $\mathcal J$ is given in (\ref{Integral-J}).
In the case of $i=j$, each term in both $A^M_{ii} \left( \mathcal W' \right)$ and $A^E_{ii} \left( \mathcal W' \right)$ is proportional to
\beq
\vert g^{\mathcal W'}_{V \, ki}  \vert^2 = \frac{g_H^2}{8} \vert \left( V^H_l \right)_{ik} \vert^2 \; ,
\eeq
which is real and contains no CP-violating phase.
The electric dipole moment $d_{l_i}(\mathcal W')$ should be vanishing 
as one can check that the sum of the two integrals in (\ref{AEWp}) vanishes when $m_i=m_j$!

We note that all our results for the charged lepton anomalous magnetic dipole moments (where $i$ and $j$ are the same charged lepton) 
are consistent with Eqs.~(3), (4), (10) and (11) in~\cite{Leveille:1977rc} if we choose $q_{\rm F}=q_{\rm X}=q_{\rm H}=-1$ in these formulas.

\renewcommand{\theequation}{B.\arabic{equation}}
\setcounter{equation}{0}
\section*{Appendix B. Expression for $\mathcal I ( m_i, m_j, m_k, m_X )$ 
in 't Hooft-Feynman gauge.}
\label{IinFeynmanGauge}

In the SM, the $W$ loop contribution can be evaluated in the 't Hooft-Feynman gauge. The longitudinal contributions from the two $W$ 
propagators will be `simulated' by three extra diagrams involving the couplings $\gamma G^+ G^-$ and 
$\gamma G^\pm W^\mp$ where $G^\pm$ are the charged Goldstone bosons of $W^\pm$. 
The expression of $\mathcal I ( m_i, m_j, m_k, m_X )$ in the 't Hooft-Feynman gauge is 
\begin{align}
\label{Integral-I-Feynman}
\mathcal I_{\, \rm ' t \,Hooft-Feynman} & \left( m_i , m_j , m_k , m_X \right) \nonumber \\ 
& = \int_0^1 {\rm d} x \int_0^{1-x} {\rm d} y \Biggl( \frac{1} { -x z m_i^2 - x y m_j^2 + x m_k^2 + (1 - x ) m_X^2 } \Biggr) \nonumber \\
& \;\;\;\; \times \Biggl\{ \Biggl[ \bigl( y + 2 z \left( 1 - x \right) \bigr) + \bigl( z + 2 y \left( 1 - x \right) \bigr) \frac{ m_j }{ m_i } - \, 3  \left( 1 - x \right)  \frac{ m_k }{ m_i }  \Biggr] 
\Biggr. \nonumber \\
& \;\;\;\;\;\;\;\;\;   - \frac{m_i^2}{m_X^2} \Biggl[ x  
\Biggl( 1 - \frac{m_k}{m_i} \Biggr) \Biggl( \frac{m_j}{m_i} - \frac{m_k}{m_i} \Biggr) 
\Biggl( z + y \frac{m_j}{m_i} + \frac{m_k}{m_i} \Biggr) \Biggr]  \nonumber \\
& \;\;\;\;\;\;\;\;\; +   \Biggl.  y \Biggl( 1 - \frac{m_k}{m_i} \Biggr)  + z  \Biggl( \frac{m_j}{m_i} - \frac{m_k}{m_i} \Biggr) \Biggl\}  \; , 
\end{align} 
where we have defined $z=1-x-y$ as before.

One can integrate over the $y$ variable in (\ref{Integral-I-Feynman}) and obtain a 1-dimension integral representation
\begin{align}
\label{Integral-I-Feynman-1D}
& \mathcal I_{\, \rm ' t \,Hooft-Feynman}  \left( m_i , m_j , m_k , m_X \right) \nonumber \\ 
& = \int_0^1 {\rm d} x \Biggl\{ A (1-x)  \Biggr. \nonumber \\
& \;\;\;\;\;\;\;\;\;  \Biggl. + \frac{B}{x} \biggl( C_0 + C_1 x + C_2 x^2 \biggr)  {\rm log}
\Biggl[ \frac{m_X^2 (1-x) - x (m_i^2 (1-x) - m_k^2) }{ m_X^2 (1-x)  - x (m_j^2 (1-x) - m_k^2) } \Biggr] \Biggr\} \; ,
\end{align}
with
\begin{align}
A & =  \frac{\left( m_i - m_k \right) \left( m_j - m_k \right) + 2 m_X^2 }{m_i \left( m_i + m_j \right) m_X^2 } \; , \\
B & = \frac{1}{m_i (m_i-m_j) (m_i + m_j)^2 m_X^2 } \; , \\
C_0 & = \left( -2 m_i^2 - 3 m_i m_j - 2 m_j^2 + 3 (m_i + m_j) m_k + m_k^2 + 2 m_X^2 \right) m_X^2 \; ,\\
C_1 & = ( m_i^2 - m_k^2 ) ( m_j^2 - m_k^2 )  + ( 2 m_i + m_j  - m_k)  ( m_i + 2 m_j  - m_k) m_X^2 - 2 m_X^4
\; ,\\
C_2 & = -m_i m_j \left( ( m_i- m_k ) (m_j -m_k)  + 2 m_X^2 \right) \; .
\end{align}
 Taking the limit of $m_j \to m_i$, (\ref{Integral-I-Feynman-1D}) reduces to 
 \begin{align}
& \mathcal I_{\, \rm ' t \,Hooft-Feynman}  \left( m_i , m_i , m_k , m_X \right) \nonumber \\ 
& = \int_0^1 {\rm d} x (1-x) \Biggl[  \frac{ 2 \left( 1-x \right) \bigl( \left( 2 - x \right) - 2 \frac{m_k}{m_i} \bigr) 
- \frac{m_i^2}{m_X^2}  x \left( 1 - \frac{m_k}{m_i} \right)^2 \left( \left(1-x \right) + \frac{m_k}{m_i} \right)  }  
{m_X^2 (1-x) - x (m_i^2 (1-x) - m_k^2 )} \Biggr] \; . 
 \end{align}
Multiplying the above result by $m_i^2 C_V^2 /8\pi^2$ and let $m_i \to m_\mu$ 
reproduces the first vector coupling piece in Eq.~ (4) (with $x \to 1-x$, $q_X = -1$ and $m_F = m_k$) of Ref.~\cite{Leveille:1977rc},
who first computed the anomaly $a_{\mu}$ for a charged $X$-loop with general gauge couplings of 
a muon and another neutral fermion $F$ in the unitary gauge. The contribution from the axial coupling can be obtained by setting $C_V \to C_A$ and flipping the 
sign of the mass $m_k$ in the above loop integral~\cite{Leveille:1977rc}. One can also reproduce the result of Ref.~\cite{Leveille:1977rc} 
by starting directly from our result~(\ref{Integral-I}) in the unitary gauge, as mentioned earlier.

An 1-dimension integral representation for $\mathcal I ( m_i, m_j, m_k, m_X )$ in the 't Hooft-Feynman gauge had been obtained 
previously in Ref.~\cite{Ma:1980gm}. Our result disagrees with this earlier result. 
One can show analytically that both of our expressions of
$\mathcal I ( m_i, m_j, m_k, m_X )$ in (\ref{Integral-I}) and (\ref{Integral-I-Feynman})  (or equivalently (\ref{Integral-I-Feynman-1D}))
from the unitary and 't Hooft-Feynman gauges respectively agree with each other. 
This can be done by integrating the integrand in (\ref{Integral-I}) over $y$ first and then subtract it with
(\ref{Integral-I-Feynman-1D}).
The difference can then be shown to be zero by applying the following identity 
\be
\int_{0}^1{\rm d}x \, g'(x) \log f(x) =  g(x) \log f(x)\bigg\vert_{0}^1 - \int_{0}^{1} {\rm d} x \frac{f'(x) g(x)}{f(x)} 
\ee
to the log terms. The intermediate steps are tedious and not illuminative, we will omit them here.

\renewcommand{\theequation}{C.\arabic{equation}}
\setcounter{equation}{0}
\section*{Appendix C. Feynman Rules}
\label{FeynRules}

Some relevant Feynman rules for this work are listed as follows.

\begin{minipage}{0.35\textwidth}
 \includegraphics[width=\textwidth]{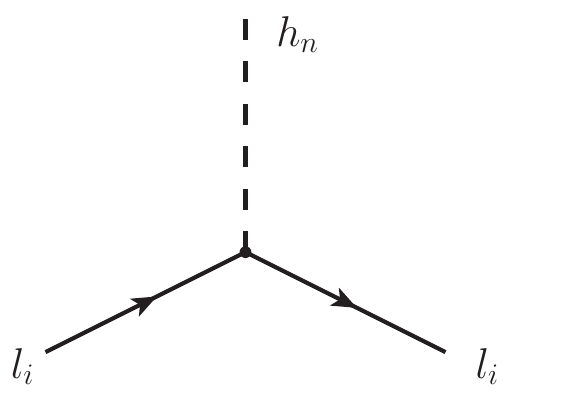}
\end{minipage}
\begin{minipage}{\textwidth}
\begin{flalign}
\hspace{2cm}
&-i(O^H)_{1n} \frac{m_i}{v}&
\end{flalign}
\end{minipage}

\begin{minipage}{0.35\textwidth}
 \includegraphics[width=\textwidth]{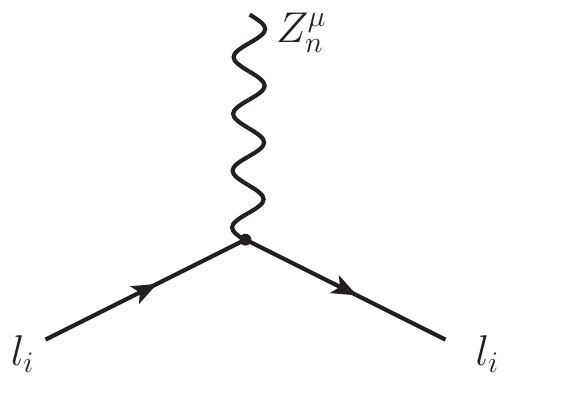}
\end{minipage}
\begin{minipage}{\textwidth}
\begin{flalign}
\hspace{2cm}
& i \gamma_\mu \left( C_{V  n} + C_{A n} \gamma_5 \right) &
\end{flalign}
\end{minipage}

\begin{minipage}{0.35\textwidth}
 \includegraphics[width=\textwidth]{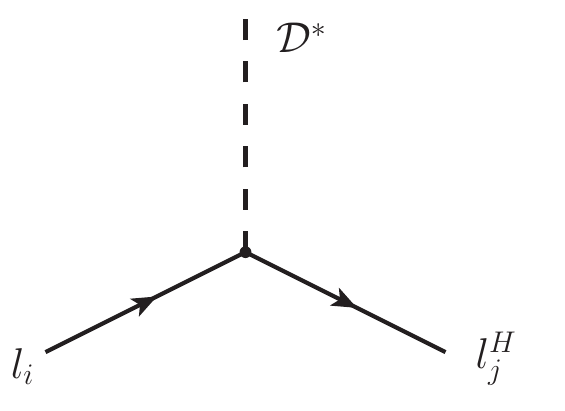}
\end{minipage}
\begin{minipage}{\textwidth}
\begin{flalign}
\hspace{2cm}
&i \left( y^{\mathcal D}_{S \, ji}   + y^{\mathcal D}_{P \, ji} \gamma_5  \right)&
\end{flalign}
\end{minipage}

\begin{minipage}{0.35\textwidth}
 \includegraphics[width=\textwidth]{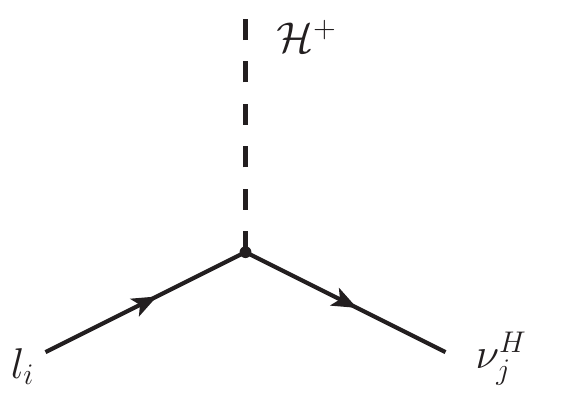}
\end{minipage}
\begin{minipage}{\textwidth}
\begin{flalign}
\hspace{2cm}
&i \left(  y^{\mathcal H}_{S \, ji}    + y^{\mathcal H}_{P \, ji}  \gamma_5 \right) &
\end{flalign}
\end{minipage}

\begin{minipage}{0.35\textwidth}
 \includegraphics[width=\textwidth]{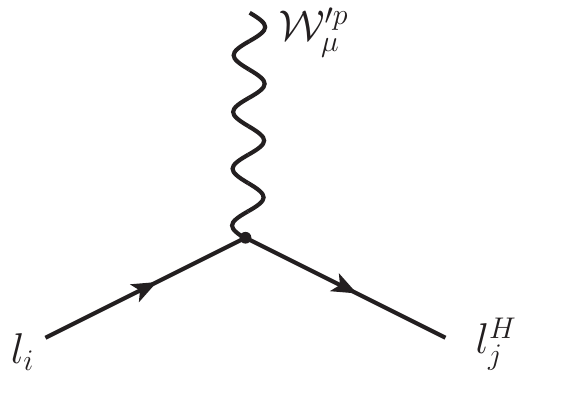}
\end{minipage}
\begin{minipage}{\textwidth}
\begin{flalign}
\hspace{2cm}
&i \gamma_\mu \left( g^{\mathcal W'}_{V \, ji}  + g^{\mathcal W'}_{A \, ji}  \, \gamma_5  \right) &
\end{flalign}
\end{minipage}

\renewcommand{\theequation}{D.\arabic{equation}}
\setcounter{equation}{0}
\section*{Appendix D. Amplitude of the on-shell $l_i \to l_j \gamma$ process}
\label{general-amplitude}
In general, the Lorentz invariant amplitude for $l_i \to l_j \gamma (i \neq j)$ as depicted in Fig. (3) is given by \cite{Cheng:1984vwu}
\be
i {\cal M}_{ji} = \left< l_j |J^{\mu}_{\rm em}|l_{i} \right>  \widetilde{\mathcal A}^{\rm ext}_\mu (q) ,
\ee
where $\widetilde{\mathcal A}^{\rm ext}_\mu (q)$  is the electromagnetic background field and  
\be
 \left< l_j |J^{\mu}_{\rm em}|l_{i} \right> = \overline{u_j}(p') \left[i \sigma^{\mu\nu} q_\nu (A + B\gamma_5) 
 + \gamma^\mu (C+D\gamma_5) + q^\mu (E + F \gamma_5) \right] u_{i}(p)
\ee
with $q=p^\prime - p$ and $A,\,B,\,C,\,D,\,E$ and $F$ are the form factors. 
Using the electromagnetic gauge invariance, one has
\be
\partial_{\mu} J^{\mu}_{\rm em} = 0
\ee 
which yields the condition 
\be
-m_i (C - D \gamma_5) + m_{j} (C + D \gamma_5) + q^2 (E + F \gamma_5) = 0
\ee
or $C = D = 0$ for  the case of $m_i \neq m_j$ and on-shell photon ($q^2 = 0$). 
Furthermore, since $q^{\mu} \widetilde{\mathcal A}^{\rm ext}_\mu = 0$, the amplitude for the on-shell $l_i \to l_j \gamma$ process is then given as 
\be
\label{matrixelement2}
i {\cal M}_{ji} =   \overline{u_j}(p') \left[ i \sigma^{\mu\nu} q_\nu (A + B\gamma_5)  \right] u_{i}(p) \widetilde{\mathcal A}^{\rm ext}_\mu(q).
\ee
To compare with the conventions established in Eq.~(\ref{matrixelement}), we can identify
\bea
A &=&-i e \frac{m_i}{2}  A_{ji}^M \, , \\
B &=& e \frac{m_i}{2}  A_{ji}^E \, . 
\eea
It is important to note that the amplitude in Eq.~(\ref{matrixelement2}) corresponds to a dimension-five operator and as such, 
it can only be induced from loop diagrams. 
Furthermore, as there can be no counterterm to absorb infinities, 
it is imperative that the amplitude in Eq.~(\ref{matrixelement2}) must be finite~\cite{Cheng:1984vwu}. 
Since the self-energy diagrams contribute only to the $C$ and $D$ form factors, they are not relevant to the on-shell amplitude of $l_i \to l_j \gamma$.
Of course these self-energy diagrams are necessarily included along with the 1PI diagrams to maintain the gauge invariance of QED!


\end{document}